%% file: FluCuNZ_HAL.tex
\documentclass[aps,prl,twocolumn,showpacs,superscriptaddress,amsmath,amssymb]{revtex4-1}

\usepackage{graphicx}
\usepackage[colorlinks,allcolors=blue]{hyperref}
\usepackage{color}

\begin{document}

\title{
Evidence for the role of proton shell closure in quasi-fission reactions from X-ray fluorescence of mass-identified fragments
}

\author{M.~Morjean}
 \email{morjean@ganil.fr}
\affiliation{
GANIL, CEA/DRF and CNRS/IN2P3, B.P. 55027, F-14076 Caen Cedex, France
}

\author{D.J.~Hinde}
\affiliation{
 Department of Nuclear Physics, Research School of Physics and Engineering, The Australian National University, ACT 0200, Australia
}

\author{C.~Simenel}
\affiliation{
 Department of Nuclear Physics, Research School of Physics and Engineering, The Australian National University, ACT 0200, Australia
}

\author{D.Y.~Jeung}
\affiliation{
 Department of Nuclear Physics, Research School of Physics and Engineering, The Australian National University, ACT 0200, Australia
}

\author{M.~Airiau}
\affiliation{
Institut de Physique Nucl\'eaire, CNRS-IN2P3, Universit\'e Paris-Sud, Universit\'e Paris-Saclay, F-91406 Orsay Cedex, France
}
\affiliation{
Irfu, CEA, Universit\'e Paris-Saclay, F-91191 Gif-sur-Yvette, France}

\author{K.J.~Cook}
\affiliation{
 Department of Nuclear Physics, Research School of Physics and Engineering, The Australian National University, ACT 0200, Australia
}

\author{M.~Dasgupta}
\affiliation{
 Department of Nuclear Physics, Research School of Physics and Engineering, The Australian National University, ACT 0200, Australia
}

\author{ A.~Drouart}
\affiliation{
Irfu, CEA, Universit\'e Paris-Saclay, F-91191 Gif-sur-Yvette, France}

\author{D.~Jacquet}
\affiliation{
Institut de Physique Nucl\'eaire, CNRS-IN2P3, Universit\'e Paris-Sud, Universit\'e Paris-Saclay, F-91406 Orsay Cedex, France
}

\author{S.~Kalkal}
\altaffiliation[Present address: ]{
 Thapar University, Patiala, Punjab 147004, India
 }
\affiliation{
 Department of Nuclear Physics, Research School of Physics and Engineering, The Australian National University, ACT 0200, Australia
}

\author{C.S.~Palshetkar}
\altaffiliation[Present address: ]{
Nuclear Physics Division, Bhabha Atomic Research Centre, Mumbai 400085, India.
}
 \affiliation{
 Department of Nuclear Physics, Research School of Physics and Engineering, The Australian National University, ACT 0200, Australia
}

\author{E.~Prasad}
\altaffiliation[Present address: ]{
Department of Physics, School of Mathematical and Physical Sciences, Central University of Kerala, Kasaragod 671314, India
}
\affiliation{
 Department of Nuclear Physics, Research School of Physics and Engineering, The Australian National University, ACT 0200, Australia}

\author{D.~Rafferty}
\affiliation{
 Department of Nuclear Physics, Research School of Physics and Engineering, The Australian National University, ACT 0200, Australia
}

\author{E.C.~Simpson}
\affiliation{
 Department of Nuclear Physics, Research School of Physics and Engineering, The Australian National University, ACT 0200, Australia
}

\author{L.~Tassan-Got}
\affiliation{
Institut de Physique Nucl\'eaire, CNRS-IN2P3, Universit\'e Paris-Sud, Universit\'e Paris-Saclay, F-91406 Orsay Cedex, France
}

\author{K.~Vo-Phuoc}
\affiliation{
 Department of Nuclear Physics, Research School of Physics and Engineering, The Australian National University, ACT 0200, Australia
}

\author{E.~Williams}
\affiliation{
 Department of Nuclear Physics, Research School of Physics and Engineering, The Australian National University, ACT 0200, Australia
}


\begin{abstract}
The atomic numbers and the masses of fragments formed in quasi-fission reactions have been simultaneously measured at scission in  $^{48}$Ti + $^{238}$U reactions at a laboratory energy of 286 MeV.
The atomic numbers were determined from measured characteristic fluorescence X-rays whereas the masses  were obtained from the emission angles and times of flight of the two emerging fragments. For the first time, thanks to this full  identification of the quasi-fission fragments on a broad angular range, the important role of the proton shell closure at Z = 82 is evidenced by the associated maximum production yield, a maximum predicted by time dependent Hartree-Fock calculations. This new experimental approach gives now access to precise studies of the time dependence of the N/Z (neutron over proton ratios of the fragments) evolution in quasi-fission reactions.  
\end{abstract}


\maketitle

Since the mid-70s, it has been known that  the formation of super-heavy nuclei by fusion is hindered by out-of-equilibrium mechanisms \cite{pet75,boc82,tok85}. In these mechanisms, the available kinetic energy can be totally dissipated and large mass transfers between the projectile and the target can occur, leading to emerging fragments quite difficult to distinguish from fragments arising from fusion followed by fission (that might be mass symmetric or asymmetric) \cite{bac85, mor08, fre12, jac09}. Due to these characteristics, the generic name quasi-fission (QF) is nowadays often used for all these mechanisms. Since the pioneering works, many experimental aspects of QF have been explored \cite{mat04,ber01,hin08,hin08b,nis12,rie13,wil13,wak14,koz16,pra16} and dynamical models, macroscopic or microscopic, have been developed in order to reproduce  cross-sections, distributions of mass, angle, kinetic or excitation energy and some of the correlations between these observables \cite{wak14,sek16,lic16,obe14,*uma15,*uma16,sim12a,sim12b,dia01,zag07,ari12}. Considering the huge experimental  difficulties to extract in a non-arbitrary way small cross-sections of fusion followed by fission from dominant quasi-fission cross-sections, a key issue for super-heavy nucleus formation studies, it is now essential to get a very good understanding of the QF mechanisms and to confront and improve the models with unambiguous exclusive data in order to reach reliable predictive capacities.  

A simultaneous determination of the fragment atomic number (Z) and mass (A) formed in QF or in fission processes remains nowadays a challenge \cite{enq01,arm04,caa13,mar15,che16}, especially difficult because these quantities are most of the time measured after particle evaporation. 
 In this letter,  an experimental approach giving access for QF fragments to A and Z at scission will be presented and the data compared with predictions of   a microscopic time dependent Hatree-Fock (TDHF) model \cite{sim12b}. 
 The atomic number was determined from the coincident characteristic fluorescence X-rays, as already attempted for fission fragments \cite{gri90}, whereas the mass was determined from the velocities of the emerging fragments.  

A $^{48}$Ti$^{19+}$ beam was accelerated at 5.75 MeV/nucleon by the Australian National University electrostatic accelerator followed by its LINAC post-accelerator. It bombarded  UF$_4$ targets highly enriched in $^{238}$U on thin carbon or aluminum backings. 
Due to damage resulting from the beam impact, the targets were rapidly drilled and different sample thicknesses, ranging from 340 up to 940 $ \mu g/cm^2$, have been used during  3  days of data acquisition with a beam intensity I $\approx 12$ nA.
For binary reactions, a very large range of folding angles between the 2 emerging fragments was covered by  2 large area position sensitive multi-wire proportional counters (280*360 mm$^2$)  MWPC1 and MWPC2. They were positioned on opposite sides of the beam at $d_1=195$ and $d_2=180$ mm from the target, covering the angular ranges $53 ^{\circ} \leqslant \theta _{1} \leqslant 124  ^ {\circ} $  and $20 ^{\circ} \leqslant \theta _{2} \leqslant 80  ^ {\circ} $, respectively.  Coincident photons were detected by three  planar germanium detectors (500 mm$^2$, 1 cm thick each) located at 6 cm from the target. These detectors were positioned at the same polar angle $\theta =143^\circ$, but at different azimuthal angles $\phi =$ 90, 330 and 210$^\circ$ with respect to the plane containing the target and the MWPC centers. 

\begin{figure}
\includegraphics[width=9cm,height=6cm]{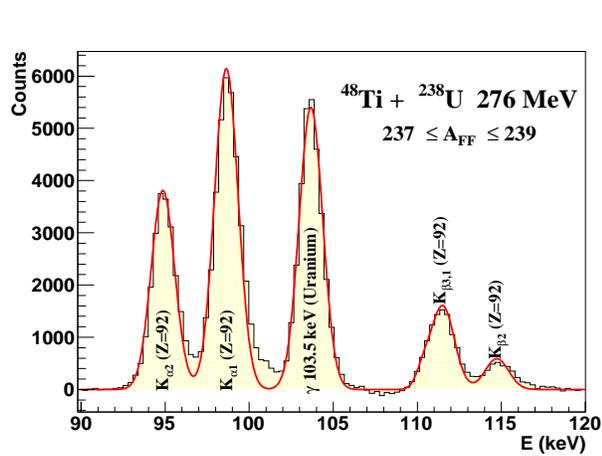}\caption{\label{fig:elast}Photon energy spectrum for fragments with A$=238\pm1$. The red line is a fit to the data (see text).}
\end{figure}

An absolute time calibration could be achieved with a precision better than 200 ps thanks to the kinematical correlation between the detection angles and the velocities for  elastically scattered projectile and target nuclei detected on broad angular ranges. A variance  $\sigma = 2.6$ amu has been inferred for the mass resolution from a dedicated measurement at 3.5 MeV/nucleon, an energy much below the Coulomb barrier in which only projectile and target nuclei could be detected. With this resolution, the average masses are quite accurately determined since,   
from the gaussian mass distributions measured at 5.75 MeV/nucleon for elastic and weakly inelastic reactions, the actual projectile and target masses are obtained with a precision better than $\pm1$ amu over the whole angular range.
 For the germanium detectors, an energy resolution $\sigma = 400$ eV is achieved  after Doppler correction in the whole energy range involved, resulting essentially  from the aperture of the germanium detectors that precludes more accurate corrections.

Figure~\ref{fig:elast} shows the photon energy spectrum for heavy fragments with A$=238\pm1$.  Uranium characteristic K X-rays and  the $\gamma$-ray at 103.5 keV from the uranium $(4^+\rightarrow2^+)$ transition  can be easily identified. The red line is a fit to the data with 6 gaussian distributions. The  centroid,  variance  and  normalization factor of each of the 6 distributions were free parameters in this fit. The 6 centroids found by the best fit differ by less than 100 eV from the tabulated energies either for the $\gamma$ transition or for K X-rays from U$^{1+}$ ions (referred to in the literature as diagram rays) \cite{xra09}.
Furthermore, the yield ratios between the different K transitions agree within less than 5$\%$ with the tabulated ones \cite{xra09}. Both these energies and yield ratios  point  out that, for the relatively low ionization states involved for the transiently formed unified atoms as well as for the emerging uranium nuclei, the relative populations on the L and M sub-shells are not sensitively modified by the processes responsible for K vacancy creation.
Therefore, since on an atomic scale elastic scattering at large angles and very central nuclear reactions correspond to the same impact parameter, the emerging QF fragments should also behave for K X-ray emission like 1$^+$ ions, as already stressed in \cite{beh79}. 

  \begin{figure}
\includegraphics[scale=0.45]{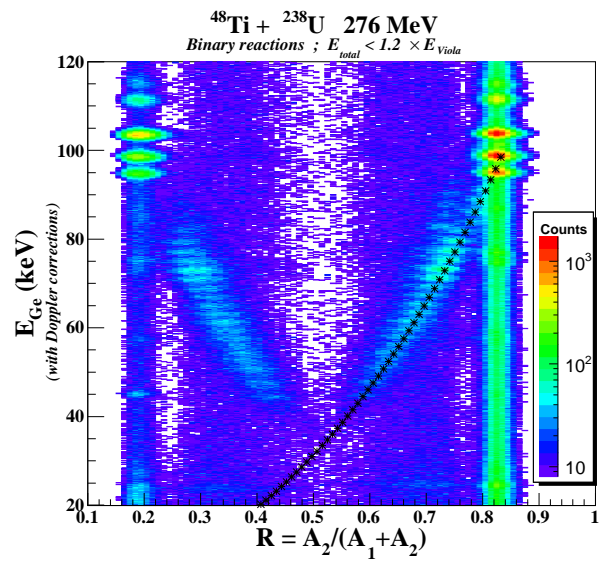}\caption{\label{fig:E_R}Photon energy versus mass ratio. The stars indicate the positions of the K$_{\alpha1}$ X-ray fluorescence lines (see text).}
\end{figure}

In the following, we shall consider only binary reactions (reactions with only 2 heavy fragments in the exit channel). Following  \cite{rie13}, binary reactions are selected from the correlation plot between the fragment velocities parallel and perpendicular to the beam axis. In addition, elastic and weakly inelastic reactions are removed by requiring total  center-of-mass kinetic energies smaller than $1.2 \times E_{Viola}$,  $E_{Viola}$ being the total energy from \cite{vio85}, adapted for asymmetric fissions. Figure~\ref{fig:E_R}  presents the correlation between the photon energy and the mass ratio R = A$_2$/(A$_1$+A$_2$), where A$_2$ (A$_1$) is the fragment mass in MWPC2 (MWPC1). The picture is not perfectly symmetric with respect to R = 0.5 due to efficiency losses for folding angles at R$ < 0.5$. The photon energies have been corrected for  Doppler shift assuming emission from the heaviest of the 2 fragments, whatever R.  Despite the selection of inelastic reactions, the different  uranium peaks already observed in Fig.~\ref{fig:elast} are still dominant for mass ratios close to those for the target or projectile nucleus (R $\approx 0.83$ or 0.17). For intermediate R values,  the Doppler correction reveals lines at constant photon energies. The stars superimposed on Fig.~\ref{fig:E_R} indicate the tabulated K$_{\alpha1}$ X-ray energies for Z$\leq92$, positioned at R values corresponding to neutron numbers  N = Z$\times\frac{\mathrm{N_{T}}} {\mathrm{Z_{T}}}$,  where the subscript T refers to the target nucleus.  The positions of the stars and of the maxima of the photon lines are quite similar, indicating thus  that the latter result essentially from K X-ray fluorescence.

\begin{figure}
\includegraphics[scale=0.45]{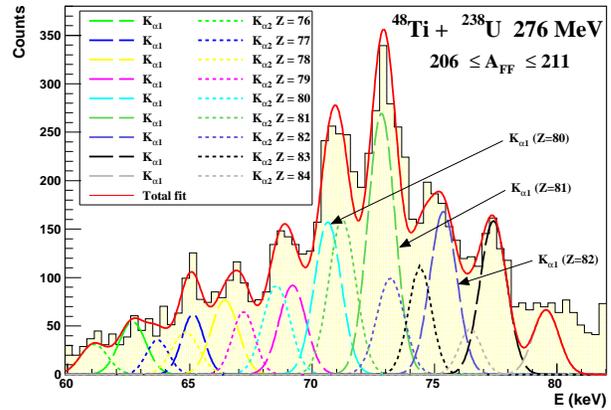}\caption{\label{fig:fit}Doppler corrected photon energy spectrum (histogram). The full red line is a fit to the data. The different contributions to this fit are shown by dotted and dashed lines (see text for details).}
\end{figure}

For the Z range covered in Fig.~\ref{fig:E_R}, K$_{\alpha1}$(Z) and K$_{\alpha2}$(Z+1) rays can have energy differences smaller than the experimental resolution and thus cannot be separated.
In order to extract the contributions of each element to the lines observed in Fig.~\ref{fig:E_R},  photon energy spectra have been constituted  for bins of fragment masses and  fits have been performed with a sum of gaussian distributions representing K$_{\alpha1}$ and  K$_{\alpha2}$ emission from all possible elements  in the energy range considered (K$_\beta$ lines contribute only weakly to the total spectra  and at energies above the K$_{\alpha}$ ones). 
Figure~\ref{fig:fit}  presents as an example the Doppler corrected spectrum after  background and random coincidence subtractions for fragments with mass $206 \leq A \leq 211$ ($0.72\leq$R$<0.74$). The red solid line shows the best fit, whereas the dashed (dotted) lines correspond for each Z to the K$_{\alpha1}$ (K$_{\alpha2}$)  contributions to this fit. 
All parameters were allowed to vary freely for fits to mass cuts with A $\geq 190$,  whereas, due to lower statistics, the ratios between the K$_{\alpha1}$ and K$_{\alpha2}$ yields had  to be fixed  for A  $< 190$ at the values for diagram lines in order to reach  satisfactory fits. 
For each Z involved,  all the best fits lead both for K$_{\alpha1}$ and K$_{\alpha2}$ lines at the tabulated energies ($\pm 0.2$ keV), with  variances $\sigma\approx 400$ eV. Furthermore, for  A $\geq 190$ they lead, within the statistical errors, to K$_{\alpha1} / $K$_{\alpha2}$ ratios in agreement with the ratios for diagram lines \cite{xra09}, as expected from the conclusions of Fig.~\ref{fig:elast} and from \cite{beh79}.  
  Therefore, as exemplified by Fig.~\ref{fig:fit}, the good behavior of all these fits as well in regions where only either a K$_{\alpha1}$ or a K$_{\alpha2}$ line is dominant as in  regions where they have similar weights  provides  us with good confidence in the yields inferred for each Z.  

The K vacancies responsible for X-ray fluorescence  from  the heavy QF atoms (with a fluorescence yield close to 1) are essentially created during the collision by direct interactions \cite{rei85} or later by internal conversion processes (IC).  Other creation mechanisms like electron shake-off are much less probable and can be neglected \cite{car68}. 
For QF reactions, the direct vacancies can be created as well in the incoming as in the outgoing part of the interaction and  their probability depends thus slightly on the emerging fragments. Nevertheless,  
the narrow mass ranges considered in the following imply narrow Z ranges and similar energy distributions, leading to similar K-vacancy creation probabilities. Therefore, the total number of direct K-vacancies is actually proportional to the number of K-vacancies present at scission. 
The very short QF lifetime (typically up to 10 zs \cite{rie13}) does not permit the decay of these vacancies that are all quasi-adiabatically  transferred to the heavier fragment, as demonstrated by molecular orbital theory \cite{rei85,anh85}. The Z distributions at scission can thus be inferred from the characteristic X-ray yields resulting from direct interactions. 
By contrast, the number of vacancies arising from IC  depends strongly on the final isotopes, after neutron evaporation, and their contributions must be subtracted from the measured X-ray yields.
IC contributions have been calculated for all the known $\gamma$- rays in NUDAT2 \cite{nudat} from all the possible nuclei after evaporation (taking into account the 2.6 amu mass resolution, and between 0 and 4 neutrons emitted). 
 Among these rays,  the ones not observed (either because they are not emitted or due to too low statistics) have all been  assumed to be Gaussian, with a maximum at the background level and  $\sigma = 400$ eV, leading  to a strong overestimation of their contributions.
From the $\gamma$-ray yields and from  the K conversion coefficients \cite{Kib08},  maximum uncertainties on the Z yields at scission have been determined.

In order to check a possible effect of instrumental electronic non-linearities on the measured masses, the  mass distributions obtained when the heavier QF fragment is detected either in MWPC2 or in MWPC1 have been compared for selections of characteristic X-rays associated with a restricted number of elements. 
 For a given photon selection, provided the electronic chains have good linearities, these 2 mass distributions must have identical maxima and widths since the X-rays are emitted by the same ions.
Differences of at most 1 amu for the maxima and similar widths have been found in the whole mass range involved, confirming therefore a precision on the mass identification better than 1 amu.

 \begin{figure}

\includegraphics[scale=0.45]{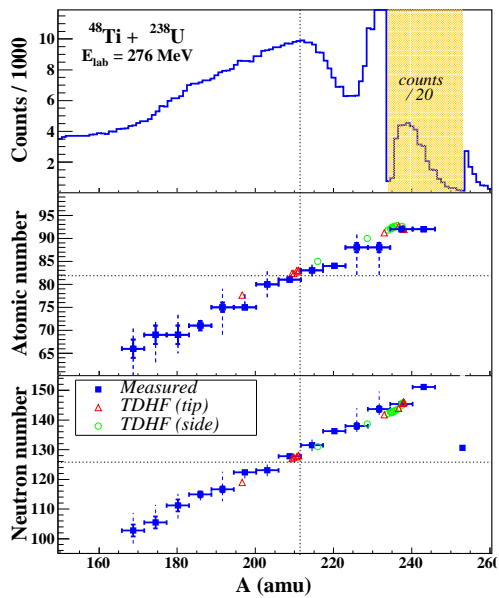}
\caption{\label{fig:AZN}
Yield, most probable atomic number and neutron number versus mass. The vertical full bars are the statistical errors. The vertical dashed lines show the maximum uncertainties resulting from internal conversion processes (see text for details).}
\end{figure}

\begin{figure}
\includegraphics[scale=0.45]{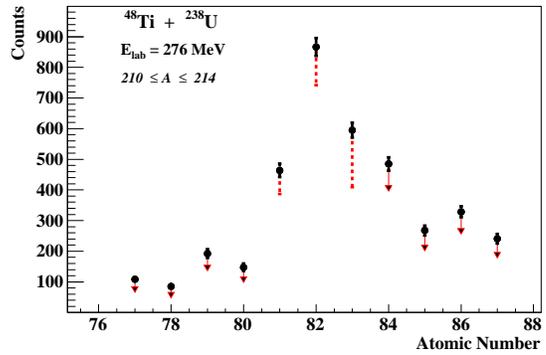}\caption{\label{fig:distZ}
 Atomic number distribution for the quasi-fission fragments with $210 \leq A \leq 214$ (see text for details).}
\end{figure}

Figure~\ref{fig:AZN}  presents the yield, the most probable atomic number at scission and the corresponding neutron number as a function of the fragment mass at scission. The horizontal full lines   show the mass range on which the most probable Z have been determined. The latter are determined with good precision in most of the cases for A$ \geq 190$, whereas, due to statistics, larger uncertainties are obtained for A $ < 190$.
As expected for QF, the maximum yield is  found for  weak mass transfers between the target and the projectile, but a secondary maximum is observed at A = 212 indicated by a dotted vertical line that crosses the data in the mid panel in the vicinity of the proton shell closure Z = 82.  By contrast, it crosses the data  in the lower panel at N $\approx 130$, and the shell closure at N = 126 would rather correspond to A $\approx$ 206, away from the maximum yield even taking into account the statistical uncertainties. Furthermore,  on the  broad range 200 $\lesssim$  A$\lesssim 224$, the influence of the proton shell closure seems to be felt in the mid panel through a lower slope. 
 The strong correlation between the maximum yield and the proton shell closure at Z = 82 is highlighted by  Fig.~\ref{fig:distZ} that presents  the Z distribution, as inferred from the fit of the  photon energy spectrum between A= 210 and 214.  The statistical errors are represented by the full vertical lines. The dashed lines indicate the maximum uncertainties resulting from IC  for Z = 81, 82 and 83. Due to the very long computational time required to take into account all the possible converted lines, IC  contributions have not been  calculated for the other Z  but, as indicated by  vertical arrows, they can only decrease the count numbers and the most probable atomic number associated with the maximum yield is found unambiguously at Z = 82. If the neutron shell closure played a dominant role, the most probable Z would be  between 84 (for A= 210) and 88 (for A=214). 
  Maximum yields around A = 208 have been already observed and analysed \cite{ari09, wak14,koz16} for QF experiments where only A was measured. They were interpreted either as arising from the N = 126 neutron shell closure \cite{koz16} or, according to TDHF calculations that do predict accumulations of fragments with Z $\approx 82$, from the proton shell closure \cite{wak14}. The simultaneous A and Z measurement provides thus a clear evidence for a dominant effect of the closed shell at Z = 82 on the QF fragment yield.  More exclusive experiments are now needed to determine if the maximum yield results from magic number influence during the primary QF fragment formation or from enhanced stability with respect to fission of primary fragments with Z $\approx 82$.  
  
  The present experimental approach can be applied to various systems,  with or without closed shells, varying the entrance channel mass and charge asymmetries,  the projectile kinetic energy, the deformations of the partners\dots The results should provide for theoretical models realistic controls of the  neutron and proton transfers between the  partners, leading thus to better predictive powers both for super-heavy synthesis by fusion and for new heavy isotope creation by transfers.
 It must be stressed that  such experiments should also provide, besides  their contribution in the super-heavy domain, quite valuable information on the symmetry energy as used in nuclear equation of state \cite{shi03,lie05,bar05}.

Triangles and circles in Fig.~\ref{fig:AZN} show the result of TDHF calculations for central collisions for tip and side target orientation (see supplemental material \cite{supmat}) and different energies (simulating for a fixed bombarding energy different angular momenta, see \cite{supmat}).  They confirm important shell effects in the $^{208}$Pb region with the tip orientation.
The neutron and proton numbers are relatively well predicted whatever the calculated mass transfers, a quantity that reflects the most probable sticking time \cite{rie13}.
However, the  number of protons (neutrons) is always slightly overestimated (underestimated), leading to N/Z values  lower than the experimental ones for the heavier QF fragments (and thus higher for the lighter fragments). 
An overestimation of the N/Z equilibration present in TDHF \cite{sim01,*sim07} might explain these differences. A similar behavior with respect to the data is obtained from  raw static energy minimizations performed as a function of the mass asymmetry but the  N/Z values for weak mass transfers are slightly lower for the heavy QF fragment than the TDHF ones, most likely due to dynamical effects in TDHF.  

 The simultaneous measurement of the atomic number and of the mass of quasi-fission fragments  over a broad angular range demonstrates the important role played in the final fragment production by the proton shell closure at Z = 82. Further experiments are now required in order to determine if the dynamical evolution of the composite system in quasi-fission is sensitive to the shell effects of each of the individual final fragments (as predicted by TDHF microscopic calculations) or if sequential fission processes favor  the survival of nuclei with magic proton numbers. The faster N/Z equilibration reached in the calculations might stress an intrinsic limitation to the TDHF approach, possibly due to the symmetry energy (and its density dependence) of the Skyrme energy density used \cite{kim97}. In the experimental approach followed for the first time here, the uncertainties arising from internal conversion processes can be greatly reduced by reasonable increases of  statistics, giving then access to the  atomic number distribution for each fragment mass.  X-ray fluorescence coupled with accurate mass determination opens thus a broad field of investigations of both quasi-fission and hot fission  processes.

The authors warmly thank N. Lobanov, T. Kib\'edi, and all the accelerator staff of the ANU Heavy Ion Accelerator for their crucial technical help.
The authors acknowledge support from the Australian Research Council through Discovery Grants No. FL110100098,  FT120100760,  DP130101569,  
DE140100784, DP160101254, and DP170102318. 
Support for accelerator operations through the NCRIS program is acknowledged. 
Two of us (CS and MA) acknowledge support from the Scientific Mobility Program of the Embassy of France in Australia.  
This research was undertaken with the assistance of resources from the National Computational Infrastructure (NCI), which is supported by the Australian Government.

\input{FluCuNZ_HAL.bbl}
\end{document}

%% file: FluCuNZ_HAL.bbl
%

%% file: FluCuNZ_HAL.bbl
\begin{thebibliography}{49}%
\makeatletter
\providecommand \@ifxundefined [1]{%
 \@ifx{#1\undefined}
}%
\providecommand \@ifnum [1]{%
 \ifnum #1\expandafter \@firstoftwo
 \else \expandafter \@secondoftwo
 \fi
}%
\providecommand \@ifx [1]{%
 \ifx #1\expandafter \@firstoftwo
 \else \expandafter \@secondoftwo
 \fi
}%
\providecommand \natexlab [1]{#1}%
\providecommand \enquote  [1]{``#1''}%
\providecommand \bibnamefont  [1]{#1}%
\providecommand \bibfnamefont [1]{#1}%
\providecommand \citenamefont [1]{#1}%
\providecommand \href@noop [0]{\@secondoftwo}%
\providecommand \href [0]{\begingroup \@sanitize@url \@href}%
\providecommand \@href[1]{\@@startlink{#1}\@@href}%
\providecommand \@@href[1]{\endgroup#1\@@endlink}%
\providecommand \@sanitize@url [0]{\catcode `\\12\catcode `\$12\catcode
  `\&12\catcode `\#12\catcode `\^12\catcode `\_12\catcode `\%12\relax}%
\providecommand \@@startlink[1]{}%
\providecommand \@@endlink[0]{}%
\providecommand \url  [0]{\begingroup\@sanitize@url \@url }%
\providecommand \@url [1]{\endgroup\@href {#1}{\urlprefix }}%
\providecommand \urlprefix  [0]{URL }%
\providecommand \Eprint [0]{\href }%
\providecommand \doibase [0]{http://dx.doi.org/}%
\providecommand \selectlanguage [0]{\@gobble}%
\providecommand \bibinfo  [0]{\@secondoftwo}%
\providecommand \bibfield  [0]{\@secondoftwo}%
\providecommand \translation [1]{[#1]}%
\providecommand \BibitemOpen [0]{}%
\providecommand \bibitemStop [0]{}%
\providecommand \bibitemNoStop [0]{.\EOS\space}%
\providecommand \EOS [0]{\spacefactor3000\relax}%
\providecommand \BibitemShut  [1]{\csname bibitem#1\endcsname}%
\let\auto@bib@innerbib\@empty
\bibitem [{\citenamefont {Peter}\ \emph {et~al.}(1975)\citenamefont {Peter},
  \citenamefont {Ngo},\ and\ \citenamefont {Tamain}}]{pet75}%
  \BibitemOpen
  \bibfield  {author} {\bibinfo {author} {\bibfnamefont {J.}~\bibnamefont
  {Peter}}, \bibinfo {author} {\bibfnamefont {C.}~\bibnamefont {Ngo}}, \ and\
  \bibinfo {author} {\bibfnamefont {B.}~\bibnamefont {Tamain}},\ }\href
  {http://search.ebscohost.com.in2p3.bib.cnrs.fr/login.aspx?direct=true&db=inh&AN=848918&lang=fr&site=ehost-live}
  {\bibfield  {journal} {\bibinfo  {journal} {Nuclear Physics A}\ }\textbf
  {\bibinfo {volume} {A250}},\ \bibinfo {pages} {351 } (\bibinfo {year}
  {1975})}\BibitemShut {NoStop}%
\bibitem [{\citenamefont {Bock}\ \emph {et~al.}(1982)\citenamefont {Bock},
  \citenamefont {Chu}, \citenamefont {Dakowski}, \citenamefont {Gobbi},
  \citenamefont {Grosse}, \citenamefont {Olmi}, \citenamefont {Sann},
  \citenamefont {Schwalm}, \citenamefont {Lynen}, \citenamefont {Muller},
  \citenamefont {Bjornholm}, \citenamefont {Esbensen}, \citenamefont {Wolfli},\
  and\ \citenamefont {Morenzoni}}]{boc82}%
  \BibitemOpen
  \bibfield  {author} {\bibinfo {author} {\bibfnamefont {R.}~\bibnamefont
  {Bock}}, \bibinfo {author} {\bibfnamefont {Y.}~\bibnamefont {Chu}}, \bibinfo
  {author} {\bibfnamefont {M.}~\bibnamefont {Dakowski}}, \bibinfo {author}
  {\bibfnamefont {A.}~\bibnamefont {Gobbi}}, \bibinfo {author} {\bibfnamefont
  {E.}~\bibnamefont {Grosse}}, \bibinfo {author} {\bibfnamefont
  {A.}~\bibnamefont {Olmi}}, \bibinfo {author} {\bibfnamefont {H.}~\bibnamefont
  {Sann}}, \bibinfo {author} {\bibfnamefont {D.}~\bibnamefont {Schwalm}},
  \bibinfo {author} {\bibfnamefont {U.}~\bibnamefont {Lynen}}, \bibinfo
  {author} {\bibfnamefont {W.}~\bibnamefont {Muller}}, \bibinfo {author}
  {\bibfnamefont {S.}~\bibnamefont {Bjornholm}}, \bibinfo {author}
  {\bibfnamefont {H.}~\bibnamefont {Esbensen}}, \bibinfo {author}
  {\bibfnamefont {W.}~\bibnamefont {Wolfli}}, \ and\ \bibinfo {author}
  {\bibfnamefont {E.}~\bibnamefont {Morenzoni}},\ }\href
  {http://search.ebscohost.com.in2p3.bib.cnrs.fr/login.aspx?direct=true&db=inh&AN=1955774&lang=fr&site=ehost-live}
  {\bibfield  {journal} {\bibinfo  {journal} {Nuclear Physics A}\ }\textbf
  {\bibinfo {volume} {A388}},\ \bibinfo {pages} {334 } (\bibinfo {year}
  {1982})}\BibitemShut {NoStop}%
\bibitem [{\citenamefont {Toke}\ \emph {et~al.}(1985)\citenamefont {Toke},
  \citenamefont {Bock}, \citenamefont {Dai}, \citenamefont {Gobbi},
  \citenamefont {Gralla}, \citenamefont {Hildenbrand}, \citenamefont
  {Kuzminski}, \citenamefont {Muller}, \citenamefont {Olmi}, \citenamefont
  {Stelzer}, \citenamefont {Back},\ and\ \citenamefont {Bjornholm}}]{tok85}%
  \BibitemOpen
  \bibfield  {author} {\bibinfo {author} {\bibfnamefont {J.}~\bibnamefont
  {Toke}}, \bibinfo {author} {\bibfnamefont {R.}~\bibnamefont {Bock}}, \bibinfo
  {author} {\bibfnamefont {G.}~\bibnamefont {Dai}}, \bibinfo {author}
  {\bibfnamefont {A.}~\bibnamefont {Gobbi}}, \bibinfo {author} {\bibfnamefont
  {S.}~\bibnamefont {Gralla}}, \bibinfo {author} {\bibfnamefont
  {K.}~\bibnamefont {Hildenbrand}}, \bibinfo {author} {\bibfnamefont
  {J.}~\bibnamefont {Kuzminski}}, \bibinfo {author} {\bibfnamefont
  {W.}~\bibnamefont {Muller}}, \bibinfo {author} {\bibfnamefont
  {A.}~\bibnamefont {Olmi}}, \bibinfo {author} {\bibfnamefont {H.}~\bibnamefont
  {Stelzer}}, \bibinfo {author} {\bibfnamefont {B.}~\bibnamefont {Back}}, \
  and\ \bibinfo {author} {\bibfnamefont {S.}~\bibnamefont {Bjornholm}},\ }\href
  {http://search.ebscohost.com.in2p3.bib.cnrs.fr/login.aspx?direct=true&db=inh&AN=2515492&lang=fr&site=ehost-live}
  {\bibfield  {journal} {\bibinfo  {journal} {Nuclear Physics A}\ }\textbf
  {\bibinfo {volume} {A440}},\ \bibinfo {pages} {327 } (\bibinfo {year}
  {1985})}\BibitemShut {NoStop}%
\bibitem [{\citenamefont {Back}(1985)}]{bac85}%
  \BibitemOpen
  \bibfield  {author} {\bibinfo {author} {\bibfnamefont {B.}~\bibnamefont
  {Back}},\ }\href
  {http://search.ebscohost.com.in2p3.bib.cnrs.fr/login.aspx?direct=true&db=inh&AN=2508106&lang=fr&site=ehost-live}
  {\bibfield  {journal} {\bibinfo  {journal} {Phys. Rev. C (Nuclear
  Physics)}\ }\textbf {\bibinfo {volume} {31}},\ \bibinfo {pages} {2104 }
  (\bibinfo {year} {1985})}\BibitemShut {NoStop}%
\bibitem [{\citenamefont {Morjean}\ \emph {et~al.}(2008)\citenamefont
  {Morjean}, \citenamefont {Jacquet}, \citenamefont {Charvet}, \citenamefont
  {L'Hoir}, \citenamefont {Laget}, \citenamefont {Parlog}, \citenamefont
  {Chbihi}, \citenamefont {Chevallier}, \citenamefont {Cohen}, \citenamefont
  {Dauvergne}, \citenamefont {Dayras}, \citenamefont {Drouart}, \citenamefont
  {Escano-Rodriguez}, \citenamefont {Frankland}, \citenamefont {Kirsch},
  \citenamefont {Lautesse}, \citenamefont {Nalpas}, \citenamefont {Ray},
  \citenamefont {Schmitt}, \citenamefont {Stodel}, \citenamefont {Tassan-Got},
  \citenamefont {Testa},\ and\ \citenamefont {Volant}}]{mor08}%
  \BibitemOpen
  \bibfield  {author} {\bibinfo {author} {\bibfnamefont {M.}~\bibnamefont
  {Morjean}}, \bibinfo {author} {\bibfnamefont {D.}~\bibnamefont {Jacquet}},
  \bibinfo {author} {\bibfnamefont {J.}~\bibnamefont {Charvet}}, \bibinfo
  {author} {\bibfnamefont {A.}~\bibnamefont {L'Hoir}}, \bibinfo {author}
  {\bibfnamefont {M.}~\bibnamefont {Laget}}, \bibinfo {author} {\bibfnamefont
  {M.}~\bibnamefont {Parlog}}, \bibinfo {author} {\bibfnamefont
  {A.}~\bibnamefont {Chbihi}}, \bibinfo {author} {\bibfnamefont
  {M.}~\bibnamefont {Chevallier}}, \bibinfo {author} {\bibfnamefont
  {C.}~\bibnamefont {Cohen}}, \bibinfo {author} {\bibfnamefont
  {D.}~\bibnamefont {Dauvergne}}, \bibinfo {author} {\bibfnamefont
  {R.}~\bibnamefont {Dayras}}, \bibinfo {author} {\bibfnamefont
  {A.}~\bibnamefont {Drouart}}, \bibinfo {author} {\bibfnamefont
  {C.}~\bibnamefont {Escano-Rodriguez}}, \bibinfo {author} {\bibfnamefont
  {J.}~\bibnamefont {Frankland}}, \bibinfo {author} {\bibfnamefont
  {R.}~\bibnamefont {Kirsch}}, \bibinfo {author} {\bibfnamefont
  {P.}~\bibnamefont {Lautesse}}, \bibinfo {author} {\bibfnamefont
  {L.}~\bibnamefont {Nalpas}}, \bibinfo {author} {\bibfnamefont
  {C.}~\bibnamefont {Ray}}, \bibinfo {author} {\bibfnamefont {C.}~\bibnamefont
  {Schmitt}}, \bibinfo {author} {\bibfnamefont {C.}~\bibnamefont {Stodel}},
  \bibinfo {author} {\bibfnamefont {L.}~\bibnamefont {Tassan-Got}}, \bibinfo
  {author} {\bibfnamefont {E.}~\bibnamefont {Testa}}, \ and\ \bibinfo {author}
  {\bibfnamefont {C.}~\bibnamefont {Volant}},\ }\href
  {http://search.ebscohost.com.in2p3.bib.cnrs.fr/login.aspx?direct=true&db=inh&AN=10158855&lang=fr&site=ehost-live}
  {\bibfield  {journal} {\bibinfo  {journal} {Physical Review Letters}\
  }\textbf {\bibinfo {volume} {101}},\ \bibinfo {pages} {072701 } (\bibinfo
  {year} {2008})}\BibitemShut {NoStop}%
\bibitem [{\citenamefont {Fr{\'e}geau}\ \emph {et~al.}(2012)\citenamefont
  {Fr{\'e}geau}, \citenamefont {Jacquet}, \citenamefont {Morjean},
  \citenamefont {Bonnet}, \citenamefont {Chbihi}, \citenamefont {Frankland},
  \citenamefont {Rivet}, \citenamefont {Tassan-Got}, \citenamefont {Dechery},
  \citenamefont {Drouart}, \citenamefont {Nalpas}, \citenamefont {Ledoux},
  \citenamefont {Parlog}, \citenamefont {Ciortea}, \citenamefont {Dumitriu},
  \citenamefont {Fluerasu}, \citenamefont {Gugiu}, \citenamefont {Gramegna},
  \citenamefont {Kravchuk}, \citenamefont {Marchi}, \citenamefont {Fabris},
  \citenamefont {Corsi},\ and\ \citenamefont {Barlini}}]{fre12}%
  \BibitemOpen
  \bibfield  {author} {\bibinfo {author} {\bibfnamefont {M.}~\bibnamefont
  {Fr{\'e}geau}}, \bibinfo {author} {\bibfnamefont {D.}~\bibnamefont
  {Jacquet}}, \bibinfo {author} {\bibfnamefont {M.}~\bibnamefont {Morjean}},
  \bibinfo {author} {\bibfnamefont {E.}~\bibnamefont {Bonnet}}, \bibinfo
  {author} {\bibfnamefont {A.}~\bibnamefont {Chbihi}}, \bibinfo {author}
  {\bibfnamefont {J.}~\bibnamefont {Frankland}}, \bibinfo {author}
  {\bibfnamefont {M.}~\bibnamefont {Rivet}}, \bibinfo {author} {\bibfnamefont
  {L.}~\bibnamefont {Tassan-Got}}, \bibinfo {author} {\bibfnamefont
  {F.}~\bibnamefont {Dechery}}, \bibinfo {author} {\bibfnamefont
  {A.}~\bibnamefont {Drouart}}, \bibinfo {author} {\bibfnamefont
  {L.}~\bibnamefont {Nalpas}}, \bibinfo {author} {\bibfnamefont
  {X.}~\bibnamefont {Ledoux}}, \bibinfo {author} {\bibfnamefont
  {M.}~\bibnamefont {Parlog}}, \bibinfo {author} {\bibfnamefont
  {C.}~\bibnamefont {Ciortea}}, \bibinfo {author} {\bibfnamefont
  {D.}~\bibnamefont {Dumitriu}}, \bibinfo {author} {\bibfnamefont
  {D.}~\bibnamefont {Fluerasu}}, \bibinfo {author} {\bibfnamefont
  {M.}~\bibnamefont {Gugiu}}, \bibinfo {author} {\bibfnamefont
  {F.}~\bibnamefont {Gramegna}}, \bibinfo {author} {\bibfnamefont
  {V.}~\bibnamefont {Kravchuk}}, \bibinfo {author} {\bibfnamefont
  {T.}~\bibnamefont {Marchi}}, \bibinfo {author} {\bibfnamefont
  {D.}~\bibnamefont {Fabris}}, \bibinfo {author} {\bibfnamefont
  {A.}~\bibnamefont {Corsi}}, \ and\ \bibinfo {author} {\bibfnamefont
  {S.}~\bibnamefont {Barlini}},\ }\href
  {http://search.ebscohost.com.in2p3.bib.cnrs.fr/login.aspx?direct=true&db=inh&AN=12667303&lang=fr&site=ehost-live}
  {\bibfield  {journal} {\bibinfo  {journal} {Physical Review Letters}\
  }\textbf {\bibinfo {volume} {108}},\ \bibinfo {pages} {122701} (\bibinfo
  {year} {2012})}\BibitemShut {NoStop}%
\bibitem [{\citenamefont {Jacquet}\ and\ \citenamefont
  {Morjean}(2009)}]{jac09}%
  \BibitemOpen
  \bibfield  {author} {\bibinfo {author} {\bibfnamefont {D.}~\bibnamefont
  {Jacquet}}\ and\ \bibinfo {author} {\bibfnamefont {M.}~\bibnamefont
  {Morjean}},\ }\href
  {http://search.ebscohost.com.in2p3.bib.cnrs.fr/login.aspx?direct=true&db=inh&AN=10945213&lang=fr&site=ehost-live}
  {\bibfield  {journal} {\bibinfo  {journal} {Progress in Particle and Nuclear
  Physics}\ }\textbf {\bibinfo {volume} {63}},\ \bibinfo {pages} {155 }
  (\bibinfo {year} {2009})}\BibitemShut {NoStop}%
\bibitem [{\citenamefont {Materna}\ \emph {et~al.}(2004)\citenamefont
  {Materna}, \citenamefont {Aritomo}, \citenamefont {Amar}, \citenamefont
  {Bogatchev}, \citenamefont {Bouchat}, \citenamefont {Dorvaux}, \citenamefont
  {Giardina}, \citenamefont {Gre´vy}, \citenamefont {Hanappe}, \citenamefont
  {Itkis}, \citenamefont {Jandel}, \citenamefont {Knyajeva}, \citenamefont
  {Kliman}, \citenamefont {Kozulin}, \citenamefont {Kondratiev}, \citenamefont
  {Krupa}, \citenamefont {Pe´ter}, \citenamefont {Prokhorova}, \citenamefont
  {Pokrovsky}, \citenamefont {Schmitt}, \citenamefont {Stuttge´},\ and\
  \citenamefont {Voskresensky}}]{mat04}%
  \BibitemOpen
  \bibfield  {author} {\bibinfo {author} {\bibfnamefont {T.}~\bibnamefont
  {Materna}}, \bibinfo {author} {\bibfnamefont {Y.}~\bibnamefont {Aritomo}},
  \bibinfo {author} {\bibfnamefont {N.}~\bibnamefont {Amar}}, \bibinfo {author}
  {\bibfnamefont {A.}~\bibnamefont {Bogatchev}}, \bibinfo {author}
  {\bibfnamefont {V.}~\bibnamefont {Bouchat}}, \bibinfo {author} {\bibfnamefont
  {O.}~\bibnamefont {Dorvaux}}, \bibinfo {author} {\bibfnamefont
  {G.}~\bibnamefont {Giardina}}, \bibinfo {author} {\bibfnamefont
  {S.}~\bibnamefont {Gre´vy}}, \bibinfo {author} {\bibfnamefont
  {F.}~\bibnamefont {Hanappe}}, \bibinfo {author} {\bibfnamefont
  {I.}~\bibnamefont {Itkis}}, \bibinfo {author} {\bibfnamefont
  {M.}~\bibnamefont {Jandel}}, \bibinfo {author} {\bibfnamefont
  {G.}~\bibnamefont {Knyajeva}}, \bibinfo {author} {\bibfnamefont
  {J.}~\bibnamefont {Kliman}}, \bibinfo {author} {\bibfnamefont
  {E.}~\bibnamefont {Kozulin}}, \bibinfo {author} {\bibfnamefont
  {N.}~\bibnamefont {Kondratiev}}, \bibinfo {author} {\bibfnamefont
  {L.}~\bibnamefont {Krupa}}, \bibinfo {author} {\bibfnamefont
  {J.}~\bibnamefont {Pe´ter}}, \bibinfo {author} {\bibfnamefont
  {E.}~\bibnamefont {Prokhorova}}, \bibinfo {author} {\bibfnamefont
  {I.}~\bibnamefont {Pokrovsky}}, \bibinfo {author} {\bibfnamefont
  {C.}~\bibnamefont {Schmitt}}, \bibinfo {author} {\bibfnamefont
  {L.}~\bibnamefont {Stuttge´}}, \ and\ \bibinfo {author} {\bibfnamefont
  {V.}~\bibnamefont {Voskresensky}},\ }\href {\doibase
  http://dx.doi.org/10.1016/j.nuclphysa.2004.01.030} {\bibfield  {journal}
  {\bibinfo  {journal} {Nuclear Physics A}\ }\textbf {\bibinfo {volume}
  {734}},\ \bibinfo {pages} {184 } (\bibinfo {year} {2004})},\ \bibinfo {note}
  {proceedings of the Eighth International Conference On Nucleus-Nucleus
  Collisions}\BibitemShut {NoStop}%
\bibitem [{\citenamefont {Berriman}\ \emph {et~al.}(2001)\citenamefont
  {Berriman}, \citenamefont {Hinde}, \citenamefont {Dasgupta}, \citenamefont
  {Morton}, \citenamefont {Butt},\ and\ \citenamefont {Newton}}]{ber01}%
  \BibitemOpen
  \bibfield  {author} {\bibinfo {author} {\bibfnamefont {A.~C.}\ \bibnamefont
  {Berriman}}, \bibinfo {author} {\bibfnamefont {D.~J.}\ \bibnamefont {Hinde}},
  \bibinfo {author} {\bibfnamefont {M.}~\bibnamefont {Dasgupta}}, \bibinfo
  {author} {\bibfnamefont {C.~R.}\ \bibnamefont {Morton}}, \bibinfo {author}
  {\bibfnamefont {R.~D.}\ \bibnamefont {Butt}}, \ and\ \bibinfo {author}
  {\bibfnamefont {J.~O.}\ \bibnamefont {Newton}},\ }\href@noop {} {\bibfield
  {journal} {\bibinfo  {journal} {Nature}\ }\textbf {\bibinfo {volume} {413}},\
  \bibinfo {pages} {144} (\bibinfo {year} {2001})}\BibitemShut {NoStop}%
\bibitem [{\citenamefont {Hinde}\ \emph
  {et~al.}(2008{\natexlab{a}})\citenamefont {Hinde}, \citenamefont {du~Rietz},
  \citenamefont {Dasgupta}, \citenamefont {Thomas},\ and\ \citenamefont
  {Gasques}}]{hin08}%
  \BibitemOpen
  \bibfield  {author} {\bibinfo {author} {\bibfnamefont {D.~J.}\ \bibnamefont
  {Hinde}}, \bibinfo {author} {\bibfnamefont {R.}~\bibnamefont {du~Rietz}},
  \bibinfo {author} {\bibfnamefont {M.}~\bibnamefont {Dasgupta}}, \bibinfo
  {author} {\bibfnamefont {R.~G.}\ \bibnamefont {Thomas}}, \ and\ \bibinfo
  {author} {\bibfnamefont {L.~R.}\ \bibnamefont {Gasques}},\ }\href {\doibase
  10.1103/PhysRevLett.101.092701} {\bibfield  {journal} {\bibinfo  {journal}
  {Phys. Rev. Lett.}\ }\textbf {\bibinfo {volume} {101}},\ \bibinfo {pages}
  {092701} (\bibinfo {year} {2008}{\natexlab{a}})}\BibitemShut {NoStop}%
\bibitem [{\citenamefont {Hinde}\ \emph
  {et~al.}(2008{\natexlab{b}})\citenamefont {Hinde}, \citenamefont {Thomas},
  \citenamefont {du~Rietz}, \citenamefont {Diaz-Torres}, \citenamefont
  {Dasgupta}, \citenamefont {Brown}, \citenamefont {Evers}, \citenamefont
  {Gasques}, \citenamefont {Rafiei},\ and\ \citenamefont {Rodriguez}}]{hin08b}%
  \BibitemOpen
  \bibfield  {author} {\bibinfo {author} {\bibfnamefont {D.~J.}\ \bibnamefont
  {Hinde}}, \bibinfo {author} {\bibfnamefont {R.~G.}\ \bibnamefont {Thomas}},
  \bibinfo {author} {\bibfnamefont {R.}~\bibnamefont {du~Rietz}}, \bibinfo
  {author} {\bibfnamefont {A.}~\bibnamefont {Diaz-Torres}}, \bibinfo {author}
  {\bibfnamefont {M.}~\bibnamefont {Dasgupta}}, \bibinfo {author}
  {\bibfnamefont {M.~L.}\ \bibnamefont {Brown}}, \bibinfo {author}
  {\bibfnamefont {M.}~\bibnamefont {Evers}}, \bibinfo {author} {\bibfnamefont
  {L.~R.}\ \bibnamefont {Gasques}}, \bibinfo {author} {\bibfnamefont
  {R.}~\bibnamefont {Rafiei}}, \ and\ \bibinfo {author} {\bibfnamefont {M.~D.}\
  \bibnamefont {Rodriguez}},\ }\href {\doibase 10.1103/PhysRevLett.100.202701}
  {\bibfield  {journal} {\bibinfo  {journal} {Phys. Rev. Lett.}\ }\textbf
  {\bibinfo {volume} {100}},\ \bibinfo {pages} {202701} (\bibinfo {year}
  {2008}{\natexlab{b}})}\BibitemShut {NoStop}%
\bibitem [{\citenamefont {Nishio}\ \emph {et~al.}(2012)\citenamefont {Nishio},
  \citenamefont {Mitsuoka}, \citenamefont {Nishinaka}, \citenamefont {Makii},
  \citenamefont {Wakabayashi}, \citenamefont {Ikezoe}, \citenamefont {Hirose},
  \citenamefont {Ohtsuki}, \citenamefont {Aritomo},\ and\ \citenamefont
  {Hofmann}}]{nis12}%
  \BibitemOpen
  \bibfield  {author} {\bibinfo {author} {\bibfnamefont {K.}~\bibnamefont
  {Nishio}}, \bibinfo {author} {\bibfnamefont {S.}~\bibnamefont {Mitsuoka}},
  \bibinfo {author} {\bibfnamefont {I.}~\bibnamefont {Nishinaka}}, \bibinfo
  {author} {\bibfnamefont {H.}~\bibnamefont {Makii}}, \bibinfo {author}
  {\bibfnamefont {Y.}~\bibnamefont {Wakabayashi}}, \bibinfo {author}
  {\bibfnamefont {H.}~\bibnamefont {Ikezoe}}, \bibinfo {author} {\bibfnamefont
  {K.}~\bibnamefont {Hirose}}, \bibinfo {author} {\bibfnamefont
  {T.}~\bibnamefont {Ohtsuki}}, \bibinfo {author} {\bibfnamefont
  {Y.}~\bibnamefont {Aritomo}}, \ and\ \bibinfo {author} {\bibfnamefont
  {S.}~\bibnamefont {Hofmann}},\ }\href {\doibase 10.1103/PhysRevC.86.034608}
  {\bibfield  {journal} {\bibinfo  {journal} {Phys. Rev. C}\ }\textbf {\bibinfo
  {volume} {86}},\ \bibinfo {pages} {034608} (\bibinfo {year}
  {2012})}\BibitemShut {NoStop}%
\bibitem [{\citenamefont {du~Rietz}\ \emph {et~al.}(2013)\citenamefont
  {du~Rietz}, \citenamefont {Williams}, \citenamefont {Hinde}, \citenamefont
  {Dasgupta}, \citenamefont {Evers}, \citenamefont {Lin}, \citenamefont
  {Luong}, \citenamefont {Simenel},\ and\ \citenamefont {Wakhle}}]{rie13}%
  \BibitemOpen
  \bibfield  {author} {\bibinfo {author} {\bibfnamefont {R.}~\bibnamefont
  {du~Rietz}}, \bibinfo {author} {\bibfnamefont {E.}~\bibnamefont {Williams}},
  \bibinfo {author} {\bibfnamefont {D.~J.}\ \bibnamefont {Hinde}}, \bibinfo
  {author} {\bibfnamefont {M.}~\bibnamefont {Dasgupta}}, \bibinfo {author}
  {\bibfnamefont {M.}~\bibnamefont {Evers}}, \bibinfo {author} {\bibfnamefont
  {C.~J.}\ \bibnamefont {Lin}}, \bibinfo {author} {\bibfnamefont {D.~H.}\
  \bibnamefont {Luong}}, \bibinfo {author} {\bibfnamefont {C.}~\bibnamefont
  {Simenel}}, \ and\ \bibinfo {author} {\bibfnamefont {A.}~\bibnamefont
  {Wakhle}},\ }\href {\doibase 10.1103/PhysRevC.88.054618} {\bibfield
  {journal} {\bibinfo  {journal} {Phys. Rev. C}\ }\textbf {\bibinfo {volume}
  {88}},\ \bibinfo {pages} {054618} (\bibinfo {year} {2013})}\BibitemShut
  {NoStop}%
\bibitem [{\citenamefont {Williams}\ \emph {et~al.}(2013)\citenamefont
  {Williams}, \citenamefont {Hinde}, \citenamefont {Dasgupta}, \citenamefont
  {du~Rietz}, \citenamefont {Carter}, \citenamefont {Evers}, \citenamefont
  {Luong}, \citenamefont {McNeil}, \citenamefont {Rafferty}, \citenamefont
  {Ramachandran},\ and\ \citenamefont {Wakhle}}]{wil13}%
  \BibitemOpen
  \bibfield  {author} {\bibinfo {author} {\bibfnamefont {E.}~\bibnamefont
  {Williams}}, \bibinfo {author} {\bibfnamefont {D.}~\bibnamefont {Hinde}},
  \bibinfo {author} {\bibfnamefont {M.}~\bibnamefont {Dasgupta}}, \bibinfo
  {author} {\bibfnamefont {R.}~\bibnamefont {du~Rietz}}, \bibinfo {author}
  {\bibfnamefont {I.}~\bibnamefont {Carter}}, \bibinfo {author} {\bibfnamefont
  {M.}~\bibnamefont {Evers}}, \bibinfo {author} {\bibfnamefont
  {D.}~\bibnamefont {Luong}}, \bibinfo {author} {\bibfnamefont
  {S.}~\bibnamefont {McNeil}}, \bibinfo {author} {\bibfnamefont
  {D.}~\bibnamefont {Rafferty}}, \bibinfo {author} {\bibfnamefont
  {K.}~\bibnamefont {Ramachandran}}, \ and\ \bibinfo {author} {\bibfnamefont
  {A.}~\bibnamefont {Wakhle}},\ }\href
  {http://search.ebscohost.com.in2p3.bib.cnrs.fr/login.aspx?direct=true&db=inh&AN=13753495&lang=fr&site=ehost-live}
  {\bibfield  {journal} {\bibinfo  {journal} {Physical Review C (Nuclear
  Physics)}\ }\textbf {\bibinfo {volume} {88}},\ \bibinfo {pages} {034611}
  (\bibinfo {year} {2013})}\BibitemShut {NoStop}%
\bibitem [{\citenamefont {Wakhle}\ \emph {et~al.}(2014)\citenamefont {Wakhle},
  \citenamefont {Simenel}, \citenamefont {Hinde}, \citenamefont {Dasgupta},
  \citenamefont {Evers}, \citenamefont {Luong}, \citenamefont {du~Rietz},\ and\
  \citenamefont {Williams}}]{wak14}%
  \BibitemOpen
  \bibfield  {author} {\bibinfo {author} {\bibfnamefont {A.}~\bibnamefont
  {Wakhle}}, \bibinfo {author} {\bibfnamefont {C.}~\bibnamefont {Simenel}},
  \bibinfo {author} {\bibfnamefont {D.~J.}\ \bibnamefont {Hinde}}, \bibinfo
  {author} {\bibfnamefont {M.}~\bibnamefont {Dasgupta}}, \bibinfo {author}
  {\bibfnamefont {M.}~\bibnamefont {Evers}}, \bibinfo {author} {\bibfnamefont
  {D.~H.}\ \bibnamefont {Luong}}, \bibinfo {author} {\bibfnamefont
  {R.}~\bibnamefont {du~Rietz}}, \ and\ \bibinfo {author} {\bibfnamefont
  {E.}~\bibnamefont {Williams}},\ }\href {\doibase
  10.1103/PhysRevLett.113.182502} {\bibfield  {journal} {\bibinfo  {journal}
  {Phys. Rev. Lett.}\ }\textbf {\bibinfo {volume} {113}},\ \bibinfo {pages}
  {182502} (\bibinfo {year} {2014})}\BibitemShut {NoStop}%
\bibitem [{\citenamefont {Kozulin}\ \emph {et~al.}(2016)\citenamefont
  {Kozulin}, \citenamefont {Knyazheva}, \citenamefont {Novikov}, \citenamefont
  {Itkis}, \citenamefont {Itkis}, \citenamefont {Dmitriev}, \citenamefont
  {Oganessian}, \citenamefont {Bogachev}, \citenamefont {Kozulina},
  \citenamefont {Harca}, \citenamefont {Trzaska},\ and\ \citenamefont
  {Ghosh}}]{koz16}%
  \BibitemOpen
  \bibfield  {author} {\bibinfo {author} {\bibfnamefont {E.~M.}\ \bibnamefont
  {Kozulin}}, \bibinfo {author} {\bibfnamefont {G.~N.}\ \bibnamefont
  {Knyazheva}}, \bibinfo {author} {\bibfnamefont {K.~V.}\ \bibnamefont
  {Novikov}}, \bibinfo {author} {\bibfnamefont {I.~M.}\ \bibnamefont {Itkis}},
  \bibinfo {author} {\bibfnamefont {M.~G.}\ \bibnamefont {Itkis}}, \bibinfo
  {author} {\bibfnamefont {S.~N.}\ \bibnamefont {Dmitriev}}, \bibinfo {author}
  {\bibfnamefont {Y.~T.}\ \bibnamefont {Oganessian}}, \bibinfo {author}
  {\bibfnamefont {A.~A.}\ \bibnamefont {Bogachev}}, \bibinfo {author}
  {\bibfnamefont {N.~I.}\ \bibnamefont {Kozulina}}, \bibinfo {author}
  {\bibfnamefont {I.}~\bibnamefont {Harca}}, \bibinfo {author} {\bibfnamefont
  {W.~H.}\ \bibnamefont {Trzaska}}, \ and\ \bibinfo {author} {\bibfnamefont
  {T.~K.}\ \bibnamefont {Ghosh}},\ }\href {\doibase 10.1103/PhysRevC.94.054613}
  {\bibfield  {journal} {\bibinfo  {journal} {Phys. Rev. C}\ }\textbf {\bibinfo
  {volume} {94}},\ \bibinfo {pages} {054613} (\bibinfo {year}
  {2016})}\BibitemShut {NoStop}%
\bibitem [{\citenamefont {Prasad}\ \emph {et~al.}(2016)\citenamefont {Prasad},
  \citenamefont {Wakhle}, \citenamefont {Hinde}, \citenamefont {Williams},
  \citenamefont {Dasgupta}, \citenamefont {Evers}, \citenamefont {Luong},
  \citenamefont {Mohanto}, \citenamefont {Simenel},\ and\ \citenamefont
  {Vo-Phuoc}}]{pra16}%
  \BibitemOpen
  \bibfield  {author} {\bibinfo {author} {\bibfnamefont {E.}~\bibnamefont
  {Prasad}}, \bibinfo {author} {\bibfnamefont {A.}~\bibnamefont {Wakhle}},
  \bibinfo {author} {\bibfnamefont {D.~J.}\ \bibnamefont {Hinde}}, \bibinfo
  {author} {\bibfnamefont {E.}~\bibnamefont {Williams}}, \bibinfo {author}
  {\bibfnamefont {M.}~\bibnamefont {Dasgupta}}, \bibinfo {author}
  {\bibfnamefont {M.}~\bibnamefont {Evers}}, \bibinfo {author} {\bibfnamefont
  {D.~H.}\ \bibnamefont {Luong}}, \bibinfo {author} {\bibfnamefont
  {G.}~\bibnamefont {Mohanto}}, \bibinfo {author} {\bibfnamefont
  {C.}~\bibnamefont {Simenel}}, \ and\ \bibinfo {author} {\bibfnamefont
  {K.}~\bibnamefont {Vo-Phuoc}},\ }\href {\doibase 10.1103/PhysRevC.93.024607}
  {\bibfield  {journal} {\bibinfo  {journal} {Phys. Rev. C}\ }\textbf {\bibinfo
  {volume} {93}},\ \bibinfo {pages} {024607} (\bibinfo {year}
  {2016})}\BibitemShut {NoStop}%
\bibitem [{\citenamefont {Sekizawa}\ and\ \citenamefont
  {Yabana}(2016)}]{sek16}%
  \BibitemOpen
  \bibfield  {author} {\bibinfo {author} {\bibfnamefont {K.}~\bibnamefont
  {Sekizawa}}\ and\ \bibinfo {author} {\bibfnamefont {K.}~\bibnamefont
  {Yabana}},\ }\href {\doibase 10.1103/PhysRevC.93.054616} {\bibfield
  {journal} {\bibinfo  {journal} {Phys. Rev. C}\ }\textbf {\bibinfo {volume}
  {93}},\ \bibinfo {pages} {054616} (\bibinfo {year} {2016})}\BibitemShut
  {NoStop}%
\bibitem [{\citenamefont {Li}\ \emph {et~al.}(2016)\citenamefont {Li},
  \citenamefont {Zhang}, \citenamefont {Li}, \citenamefont {Zhu}, \citenamefont
  {Tian}, \citenamefont {Wang},\ and\ \citenamefont {Zhang}}]{lic16}%
  \BibitemOpen
  \bibfield  {author} {\bibinfo {author} {\bibfnamefont {C.}~\bibnamefont
  {Li}}, \bibinfo {author} {\bibfnamefont {F.}~\bibnamefont {Zhang}}, \bibinfo
  {author} {\bibfnamefont {J.}~\bibnamefont {Li}}, \bibinfo {author}
  {\bibfnamefont {L.}~\bibnamefont {Zhu}}, \bibinfo {author} {\bibfnamefont
  {J.}~\bibnamefont {Tian}}, \bibinfo {author} {\bibfnamefont {N.}~\bibnamefont
  {Wang}}, \ and\ \bibinfo {author} {\bibfnamefont {F.-S.}\ \bibnamefont
  {Zhang}},\ }\href {\doibase 10.1103/PhysRevC.93.014618} {\bibfield  {journal}
  {\bibinfo  {journal} {Phys. Rev. C}\ }\textbf {\bibinfo {volume} {93}},\
  \bibinfo {pages} {014618} (\bibinfo {year} {2016})}\BibitemShut {NoStop}%
\bibitem [{\citenamefont {Oberacker}\ \emph {et~al.}(2014)\citenamefont
  {Oberacker}, \citenamefont {Umar},\ and\ \citenamefont {Simenel}}]{obe14}%
  \BibitemOpen
  \bibfield  {author} {\bibinfo {author} {\bibfnamefont {V.~E.}\ \bibnamefont
  {Oberacker}}, \bibinfo {author} {\bibfnamefont {A.~S.}\ \bibnamefont {Umar}},
  \ and\ \bibinfo {author} {\bibfnamefont {C.}~\bibnamefont {Simenel}},\ }\href
  {\doibase 10.1103/PhysRevC.90.054605} {\bibfield  {journal} {\bibinfo
  {journal} {Phys. Rev. C}\ }\textbf {\bibinfo {volume} {90}},\ \bibinfo
  {pages} {054605} (\bibinfo {year} {2014})}\BibitemShut {NoStop}%
\bibitem [{\citenamefont {Umar}\ \emph {et~al.}(2015)\citenamefont {Umar},
  \citenamefont {Oberacker},\ and\ \citenamefont {Simenel}}]{uma15}%
  \BibitemOpen
  \bibfield  {author} {\bibinfo {author} {\bibfnamefont {A.}~\bibnamefont
  {Umar}}, \bibinfo {author} {\bibfnamefont {V.}~\bibnamefont {Oberacker}}, \
  and\ \bibinfo {author} {\bibfnamefont {C.}~\bibnamefont {Simenel}},\ }\href
  {http://search.ebscohost.com.in2p3.bib.cnrs.fr/login.aspx?direct=true&db=inh&AN=15397046&lang=fr&site=ehost-live}
  {\bibfield  {journal} {\bibinfo  {journal} {Physical Review C (Nuclear
  Physics)}\ }\textbf {\bibinfo {volume} {92}},\ \bibinfo {pages} {024621}
  (\bibinfo {year} {2015})}\BibitemShut {NoStop}%
\bibitem [{\citenamefont {Umar}\ \emph {et~al.}(2016)\citenamefont {Umar},
  \citenamefont {Oberacker},\ and\ \citenamefont {Simenel}}]{uma16}%
  \BibitemOpen
  \bibfield  {author} {\bibinfo {author} {\bibfnamefont {A.}~\bibnamefont
  {Umar}}, \bibinfo {author} {\bibfnamefont {V.}~\bibnamefont {Oberacker}}, \
  and\ \bibinfo {author} {\bibfnamefont {C.}~\bibnamefont {Simenel}},\ }\href
  {http://search.ebscohost.com.in2p3.bib.cnrs.fr/login.aspx?direct=true&db=inh&AN=16190570&lang=fr&site=ehost-live}
  {\bibfield  {journal} {\bibinfo  {journal} {Physical Review C}\ }\textbf
  {\bibinfo {volume} {94}},\ \bibinfo {pages} {024605} (\bibinfo {year}
  {2016})}\BibitemShut {NoStop}%
\bibitem [{\citenamefont {Simenel}\ \emph {et~al.}(2012)\citenamefont
  {Simenel}, \citenamefont {Hinde}, \citenamefont {du~Rietz}, \citenamefont
  {Dasgupta}, \citenamefont {Evers}, \citenamefont {Lin}, \citenamefont
  {Luong},\ and\ \citenamefont {Wakhle}}]{sim12a}%
  \BibitemOpen
  \bibfield  {author} {\bibinfo {author} {\bibfnamefont {C.}~\bibnamefont
  {Simenel}}, \bibinfo {author} {\bibfnamefont {D.}~\bibnamefont {Hinde}},
  \bibinfo {author} {\bibfnamefont {R.}~\bibnamefont {du~Rietz}}, \bibinfo
  {author} {\bibfnamefont {M.}~\bibnamefont {Dasgupta}}, \bibinfo {author}
  {\bibfnamefont {M.}~\bibnamefont {Evers}}, \bibinfo {author} {\bibfnamefont
  {C.}~\bibnamefont {Lin}}, \bibinfo {author} {\bibfnamefont {D.}~\bibnamefont
  {Luong}}, \ and\ \bibinfo {author} {\bibfnamefont {A.}~\bibnamefont
  {Wakhle}},\ }\href {\doibase
  http://dx.doi.org/10.1016/j.physletb.2012.03.063} {\bibfield  {journal}
  {\bibinfo  {journal} {Physics Letters B}\ }\textbf {\bibinfo {volume}
  {710}},\ \bibinfo {pages} {607 } (\bibinfo {year} {2012})}\BibitemShut
  {NoStop}%
\bibitem [{\citenamefont {Simenel}(2012)}]{sim12b}%
  \BibitemOpen
  \bibfield  {author} {\bibinfo {author} {\bibfnamefont {C.}~\bibnamefont
  {Simenel}},\ }\href {\doibase 10.1140/epja/i2012-12152-0} {\bibfield
  {journal} {\bibinfo  {journal} {The European Physical Journal A}\ }\textbf
  {\bibinfo {volume} {48}},\ \bibinfo {pages} {152} (\bibinfo {year}
  {2012})}\BibitemShut {NoStop}%
\bibitem [{\citenamefont {Diaz-Torres}\ \emph {et~al.}(2001)\citenamefont
  {Diaz-Torres}, \citenamefont {Adamian}, \citenamefont {Antonenko},\ and\
  \citenamefont {Scheid}}]{dia01}%
  \BibitemOpen
  \bibfield  {author} {\bibinfo {author} {\bibfnamefont {A.}~\bibnamefont
  {Diaz-Torres}}, \bibinfo {author} {\bibfnamefont {G.~G.}\ \bibnamefont
  {Adamian}}, \bibinfo {author} {\bibfnamefont {N.~V.}\ \bibnamefont
  {Antonenko}}, \ and\ \bibinfo {author} {\bibfnamefont {W.}~\bibnamefont
  {Scheid}},\ }\href {\doibase 10.1103/PhysRevC.64.024604} {\bibfield
  {journal} {\bibinfo  {journal} {Phys. Rev. C}\ }\textbf {\bibinfo {volume}
  {64}},\ \bibinfo {pages} {024604} (\bibinfo {year} {2001})}\BibitemShut
  {NoStop}%
\bibitem [{\citenamefont {Zagrebaev}\ and\ \citenamefont
  {Greiner}(2007)}]{zag07}%
  \BibitemOpen
  \bibfield  {author} {\bibinfo {author} {\bibfnamefont {V.}~\bibnamefont
  {Zagrebaev}}\ and\ \bibinfo {author} {\bibfnamefont {W.}~\bibnamefont
  {Greiner}},\ }\href {http://stacks.iop.org/0954-3899/34/i=1/a=001} {\bibfield
   {journal} {\bibinfo  {journal} {J. Phys. G: Nuclear and Particle
  Physics}\ }\textbf {\bibinfo {volume} {34}},\ \bibinfo {pages} {1} (\bibinfo
  {year} {2007})}\BibitemShut {NoStop}%
\bibitem [{\citenamefont {Aritomo}\ \emph {et~al.}(2012)\citenamefont
  {Aritomo}, \citenamefont {Hagino}, \citenamefont {Nishio},\ and\
  \citenamefont {Chiba}}]{ari12}%
  \BibitemOpen
  \bibfield  {author} {\bibinfo {author} {\bibfnamefont {Y.}~\bibnamefont
  {Aritomo}}, \bibinfo {author} {\bibfnamefont {K.}~\bibnamefont {Hagino}},
  \bibinfo {author} {\bibfnamefont {K.}~\bibnamefont {Nishio}}, \ and\ \bibinfo
  {author} {\bibfnamefont {S.}~\bibnamefont {Chiba}},\ }\href {\doibase
  10.1103/PhysRevC.85.044614} {\bibfield  {journal} {\bibinfo  {journal} {Phys.
  Rev. C}\ }\textbf {\bibinfo {volume} {85}},\ \bibinfo {pages} {044614}
  (\bibinfo {year} {2012})}\BibitemShut {NoStop}%
\bibitem [{\citenamefont {Enqvist}\ \emph {et~al.}(2001)\citenamefont
  {Enqvist}, \citenamefont {Wlaz{\l}o}, \citenamefont {Armbruster},
  \citenamefont {Benlliure}, \citenamefont {Bernas}, \citenamefont {Boudard},
  \citenamefont {Czajkowski}, \citenamefont {Legrain}, \citenamefont {Leray},
  \citenamefont {Mustapha}, \citenamefont {Pravikoff}, \citenamefont {Rejmund},
  \citenamefont {Schmidt}, \citenamefont {St{\'e}phan}, \citenamefont {Taieb},
  \citenamefont {Tassan-Got},\ and\ \citenamefont {Volant}}]{enq01}%
  \BibitemOpen
  \bibfield  {author} {\bibinfo {author} {\bibfnamefont {T.}~\bibnamefont
  {Enqvist}}, \bibinfo {author} {\bibfnamefont {W.}~\bibnamefont {Wlaz{\l}o}},
  \bibinfo {author} {\bibfnamefont {P.}~\bibnamefont {Armbruster}}, \bibinfo
  {author} {\bibfnamefont {J.}~\bibnamefont {Benlliure}}, \bibinfo {author}
  {\bibfnamefont {M.}~\bibnamefont {Bernas}}, \bibinfo {author} {\bibfnamefont
  {A.}~\bibnamefont {Boudard}}, \bibinfo {author} {\bibfnamefont
  {S.}~\bibnamefont {Czajkowski}}, \bibinfo {author} {\bibfnamefont
  {R.}~\bibnamefont {Legrain}}, \bibinfo {author} {\bibfnamefont
  {S.}~\bibnamefont {Leray}}, \bibinfo {author} {\bibfnamefont
  {B.}~\bibnamefont {Mustapha}}, \bibinfo {author} {\bibfnamefont
  {M.}~\bibnamefont {Pravikoff}}, \bibinfo {author} {\bibfnamefont
  {F.}~\bibnamefont {Rejmund}}, \bibinfo {author} {\bibfnamefont {K.-H.}\
  \bibnamefont {Schmidt}}, \bibinfo {author} {\bibfnamefont {C.}~\bibnamefont
  {St{\'e}phan}}, \bibinfo {author} {\bibfnamefont {J.}~\bibnamefont {Taieb}},
  \bibinfo {author} {\bibfnamefont {L.}~\bibnamefont {Tassan-Got}}, \ and\
  \bibinfo {author} {\bibfnamefont {C.}~\bibnamefont {Volant}},\ }\href
  {\doibase http://dx.doi.org/10.1016/S0375-9474(00)00563-7} {\bibfield
  {journal} {\bibinfo  {journal} {Nuclear Physics A}\ }\textbf {\bibinfo
  {volume} {686}},\ \bibinfo {pages} {481 } (\bibinfo {year}
  {2001})}\BibitemShut {NoStop}%
\bibitem [{\citenamefont {Armbruster}\ \emph {et~al.}(2004)\citenamefont
  {Armbruster}, \citenamefont {Benlliure}, \citenamefont {Bernas},
  \citenamefont {Boudard}, \citenamefont {Casarejos}, \citenamefont
  {Czajkowski}, \citenamefont {Enqvist}, \citenamefont {Leray}, \citenamefont
  {Napolitani}, \citenamefont {Pereira}, \citenamefont {Rejmund}, \citenamefont
  {Ricciardi}, \citenamefont {Schmidt}, \citenamefont {St\'ephan},
  \citenamefont {Taieb}, \citenamefont {Tassan-Got},\ and\ \citenamefont
  {Volant}}]{arm04}%
  \BibitemOpen
  \bibfield  {author} {\bibinfo {author} {\bibfnamefont {P.}~\bibnamefont
  {Armbruster}}, \bibinfo {author} {\bibfnamefont {J.}~\bibnamefont
  {Benlliure}}, \bibinfo {author} {\bibfnamefont {M.}~\bibnamefont {Bernas}},
  \bibinfo {author} {\bibfnamefont {A.}~\bibnamefont {Boudard}}, \bibinfo
  {author} {\bibfnamefont {E.}~\bibnamefont {Casarejos}}, \bibinfo {author}
  {\bibfnamefont {S.}~\bibnamefont {Czajkowski}}, \bibinfo {author}
  {\bibfnamefont {T.}~\bibnamefont {Enqvist}}, \bibinfo {author} {\bibfnamefont
  {S.}~\bibnamefont {Leray}}, \bibinfo {author} {\bibfnamefont
  {P.}~\bibnamefont {Napolitani}}, \bibinfo {author} {\bibfnamefont
  {J.}~\bibnamefont {Pereira}}, \bibinfo {author} {\bibfnamefont
  {F.}~\bibnamefont {Rejmund}}, \bibinfo {author} {\bibfnamefont {M.-V.}\
  \bibnamefont {Ricciardi}}, \bibinfo {author} {\bibfnamefont {K.-H.}\
  \bibnamefont {Schmidt}}, \bibinfo {author} {\bibfnamefont {C.}~\bibnamefont
  {St\'ephan}}, \bibinfo {author} {\bibfnamefont {J.}~\bibnamefont {Taieb}},
  \bibinfo {author} {\bibfnamefont {L.}~\bibnamefont {Tassan-Got}}, \ and\
  \bibinfo {author} {\bibfnamefont {C.}~\bibnamefont {Volant}},\ }\href
  {\doibase 10.1103/PhysRevLett.93.212701} {\bibfield  {journal} {\bibinfo
  {journal} {Phys. Rev. Lett.}\ }\textbf {\bibinfo {volume} {93}},\ \bibinfo
  {pages} {212701} (\bibinfo {year} {2004})}\BibitemShut {NoStop}%
\bibitem [{\citenamefont {Caama\~no}\ \emph {et~al.}(2013)\citenamefont
  {Caama\~no}, \citenamefont {Delaune}, \citenamefont {Farget}, \citenamefont
  {Derkx}, \citenamefont {Schmidt}, \citenamefont {Audouin}, \citenamefont
  {Bacri}, \citenamefont {Barreau}, \citenamefont {Benlliure}, \citenamefont
  {Casarejos}, \citenamefont {Chbihi}, \citenamefont
  {Fern\'andez-Dom\'{\i}nguez}, \citenamefont {Gaudefroy}, \citenamefont
  {Golabek}, \citenamefont {Jurado}, \citenamefont {Lemasson}, \citenamefont
  {Navin}, \citenamefont {Rejmund}, \citenamefont {Roger}, \citenamefont
  {Shrivastava},\ and\ \citenamefont {Schmitt}}]{caa13}%
  \BibitemOpen
  \bibfield  {author} {\bibinfo {author} {\bibfnamefont {M.}~\bibnamefont
  {Caama\~no}}, \bibinfo {author} {\bibfnamefont {O.}~\bibnamefont {Delaune}},
  \bibinfo {author} {\bibfnamefont {F.}~\bibnamefont {Farget}}, \bibinfo
  {author} {\bibfnamefont {X.}~\bibnamefont {Derkx}}, \bibinfo {author}
  {\bibfnamefont {K.-H.}\ \bibnamefont {Schmidt}}, \bibinfo {author}
  {\bibfnamefont {L.}~\bibnamefont {Audouin}}, \bibinfo {author} {\bibfnamefont
  {C.-O.}\ \bibnamefont {Bacri}}, \bibinfo {author} {\bibfnamefont
  {G.}~\bibnamefont {Barreau}}, \bibinfo {author} {\bibfnamefont
  {J.}~\bibnamefont {Benlliure}}, \bibinfo {author} {\bibfnamefont
  {E.}~\bibnamefont {Casarejos}}, \bibinfo {author} {\bibfnamefont
  {A.}~\bibnamefont {Chbihi}}, \bibinfo {author} {\bibfnamefont
  {B.}~\bibnamefont {Fern\'andez-Dom\'{\i}nguez}}, \bibinfo {author}
  {\bibfnamefont {L.}~\bibnamefont {Gaudefroy}}, \bibinfo {author}
  {\bibfnamefont {C.}~\bibnamefont {Golabek}}, \bibinfo {author} {\bibfnamefont
  {B.}~\bibnamefont {Jurado}}, \bibinfo {author} {\bibfnamefont
  {A.}~\bibnamefont {Lemasson}}, \bibinfo {author} {\bibfnamefont
  {A.}~\bibnamefont {Navin}}, \bibinfo {author} {\bibfnamefont
  {M.}~\bibnamefont {Rejmund}}, \bibinfo {author} {\bibfnamefont
  {T.}~\bibnamefont {Roger}}, \bibinfo {author} {\bibfnamefont
  {A.}~\bibnamefont {Shrivastava}}, \ and\ \bibinfo {author} {\bibfnamefont
  {C.}~\bibnamefont {Schmitt}},\ }\href {\doibase 10.1103/PhysRevC.88.024605}
  {\bibfield  {journal} {\bibinfo  {journal} {Phys. Rev. C}\ }\textbf {\bibinfo
  {volume} {88}},\ \bibinfo {pages} {024605} (\bibinfo {year}
  {2013})}\BibitemShut {NoStop}%
\bibitem [{\citenamefont {Martin}\ \emph {et~al.}(2015)\citenamefont {Martin},
  \citenamefont {Taieb}, \citenamefont {Chatillon}, \citenamefont {B{\'e}lier},
  \citenamefont {Boutoux}, \citenamefont {Ebran}, \citenamefont {Gorbinet},
  \citenamefont {Grente}, \citenamefont {Laurent}, \citenamefont {Pellereau},
  \citenamefont {Alvarez-Pol}, \citenamefont {Audouin}, \citenamefont {Aumann},
  \citenamefont {Ayyad}, \citenamefont {Benlliure}, \citenamefont {Casarejos},
  \citenamefont {Cortina~Gil}, \citenamefont {Caama{\~n}o}, \citenamefont
  {Farget}, \citenamefont {Fernandez~Dominguez}, \citenamefont {Heinz},
  \citenamefont {Jurado}, \citenamefont {Kelic-Heil}, \citenamefont {Kurz},
  \citenamefont {Nociforo}, \citenamefont {Paradela}, \citenamefont {Pietri},
  \citenamefont {Ramos}, \citenamefont {Rodriguez-Sanchez}, \citenamefont
  {Rodriguez-Tajes}, \citenamefont {Rossi}, \citenamefont {Schmidt},
  \citenamefont {Simon}, \citenamefont {Tassan-Got}, \citenamefont {Vargas},
  \citenamefont {Voss},\ and\ \citenamefont {Weick}}]{mar15}%
  \BibitemOpen
  \bibfield  {author} {\bibinfo {author} {\bibfnamefont {J.-F.}\ \bibnamefont
  {Martin}}, \bibinfo {author} {\bibfnamefont {J.}~\bibnamefont {Taieb}},
  \bibinfo {author} {\bibfnamefont {A.}~\bibnamefont {Chatillon}}, \bibinfo
  {author} {\bibfnamefont {G.}~\bibnamefont {B{\'e}lier}}, \bibinfo {author}
  {\bibfnamefont {G.}~\bibnamefont {Boutoux}}, \bibinfo {author} {\bibfnamefont
  {A.}~\bibnamefont {Ebran}}, \bibinfo {author} {\bibfnamefont
  {T.}~\bibnamefont {Gorbinet}}, \bibinfo {author} {\bibfnamefont
  {L.}~\bibnamefont {Grente}}, \bibinfo {author} {\bibfnamefont
  {B.}~\bibnamefont {Laurent}}, \bibinfo {author} {\bibfnamefont
  {E.}~\bibnamefont {Pellereau}}, \bibinfo {author} {\bibfnamefont
  {H.}~\bibnamefont {Alvarez-Pol}}, \bibinfo {author} {\bibfnamefont
  {L.}~\bibnamefont {Audouin}}, \bibinfo {author} {\bibfnamefont
  {T.}~\bibnamefont {Aumann}}, \bibinfo {author} {\bibfnamefont
  {Y.}~\bibnamefont {Ayyad}}, \bibinfo {author} {\bibfnamefont
  {J.}~\bibnamefont {Benlliure}}, \bibinfo {author} {\bibfnamefont
  {E.}~\bibnamefont {Casarejos}}, \bibinfo {author} {\bibfnamefont
  {D.}~\bibnamefont {Cortina~Gil}}, \bibinfo {author} {\bibfnamefont
  {M.}~\bibnamefont {Caama{\~n}o}}, \bibinfo {author} {\bibfnamefont
  {F.}~\bibnamefont {Farget}}, \bibinfo {author} {\bibfnamefont
  {B.}~\bibnamefont {Fernandez~Dominguez}}, \bibinfo {author} {\bibfnamefont
  {A.}~\bibnamefont {Heinz}}, \bibinfo {author} {\bibfnamefont
  {B.}~\bibnamefont {Jurado}}, \bibinfo {author} {\bibfnamefont
  {A.}~\bibnamefont {Kelic-Heil}}, \bibinfo {author} {\bibfnamefont
  {N.}~\bibnamefont {Kurz}}, \bibinfo {author} {\bibfnamefont {C.}~\bibnamefont
  {Nociforo}}, \bibinfo {author} {\bibfnamefont {C.}~\bibnamefont {Paradela}},
  \bibinfo {author} {\bibfnamefont {S.}~\bibnamefont {Pietri}}, \bibinfo
  {author} {\bibfnamefont {D.}~\bibnamefont {Ramos}}, \bibinfo {author}
  {\bibfnamefont {J.-L.}\ \bibnamefont {Rodriguez-Sanchez}}, \bibinfo {author}
  {\bibfnamefont {C.}~\bibnamefont {Rodriguez-Tajes}}, \bibinfo {author}
  {\bibfnamefont {D.}~\bibnamefont {Rossi}}, \bibinfo {author} {\bibfnamefont
  {K.-H.}\ \bibnamefont {Schmidt}}, \bibinfo {author} {\bibfnamefont
  {H.}~\bibnamefont {Simon}}, \bibinfo {author} {\bibfnamefont
  {L.}~\bibnamefont {Tassan-Got}}, \bibinfo {author} {\bibfnamefont
  {J.}~\bibnamefont {Vargas}}, \bibinfo {author} {\bibfnamefont
  {B.}~\bibnamefont {Voss}}, \ and\ \bibinfo {author} {\bibfnamefont
  {H.}~\bibnamefont {Weick}},\ }\href
  {http://search.ebscohost.com.in2p3.bib.cnrs.fr/login.aspx?direct=true&db=inh&AN=15834082&lang=fr&site=ehost-live}
  {\bibfield  {journal} {\bibinfo  {journal} {EPJ A -
  Hadrons and Nuclei}\ }\textbf {\bibinfo {volume} {51}},\ \bibinfo {pages}
  {174} (\bibinfo {year} {2015})}\BibitemShut {NoStop}%
\bibitem [{\citenamefont {Chebboubi}\ \emph {et~al.}(2016)\citenamefont
  {Chebboubi}, \citenamefont {Kessedjian}, \citenamefont {Faust}, \citenamefont
  {Blanc}, \citenamefont {Jentschel}, \citenamefont {K{\"o}ster}, \citenamefont
  {Materna}, \citenamefont {M{\'e}plan}, \citenamefont {Sage},\ and\
  \citenamefont {Serot}}]{che16}%
  \BibitemOpen
  \bibfield  {author} {\bibinfo {author} {\bibfnamefont {A.}~\bibnamefont
  {Chebboubi}}, \bibinfo {author} {\bibfnamefont {G.}~\bibnamefont
  {Kessedjian}}, \bibinfo {author} {\bibfnamefont {H.}~\bibnamefont {Faust}},
  \bibinfo {author} {\bibfnamefont {A.}~\bibnamefont {Blanc}}, \bibinfo
  {author} {\bibfnamefont {M.}~\bibnamefont {Jentschel}}, \bibinfo {author}
  {\bibfnamefont {U.}~\bibnamefont {K{\"o}ster}}, \bibinfo {author}
  {\bibfnamefont {T.}~\bibnamefont {Materna}}, \bibinfo {author} {\bibfnamefont
  {O.}~\bibnamefont {M{\'e}plan}}, \bibinfo {author} {\bibfnamefont
  {C.}~\bibnamefont {Sage}}, \ and\ \bibinfo {author} {\bibfnamefont
  {O.}~\bibnamefont {Serot}},\ }\href {\doibase
  http://dx.doi.org/10.1016/j.nimb.2016.02.008} {\bibfield  {journal} {\bibinfo
   {journal} {NIM B}\ }\textbf {\bibinfo {volume}
  {376}},\ \bibinfo {pages} {120 } (\bibinfo {year} {2016})}\BibitemShut {NoStop}%
\bibitem [{\citenamefont {Griffin}(1990)}]{gri90}%
  \BibitemOpen
  \bibfield  {author} {\bibinfo {author} {\bibfnamefont {H.}~\bibnamefont
  {Griffin}},\ }\href
  {http://search.ebscohost.com.in2p3.bib.cnrs.fr/login.aspx?direct=true&db=inh&AN=3825022&lang=fr&site=ehost-live}
  {\bibfield  {journal} {\bibinfo  {journal} {J. Rad.
  Nucl. Chem.}\ }\textbf {\bibinfo {volume} {142}},\ \bibinfo
  {pages} {279 } (\bibinfo {year} {1990})}\BibitemShut {NoStop}%
\bibitem [{\citenamefont {Kortright}\ and\ \citenamefont
  {Thomson}(2009)}]{xra09}%
  \BibitemOpen
  \bibfield  {author} {\bibinfo {author} {\bibfnamefont {J.}~\bibnamefont
  {Kortright}}\ and\ \bibinfo {author} {\bibfnamefont {A.}~\bibnamefont
  {Thomson}},\ }\href@noop {} {\emph {\bibinfo {title} {X-ray data booklet}}},\
  edited by\ \bibinfo {editor} {\bibfnamefont {A.}~\bibnamefont {Thomson}},\
  \bibinfo {number} {1-8}\ (\bibinfo  {publisher} {Lawrence Berkeley National
  Laboratory},\ \bibinfo {year} {2009})\BibitemShut {NoStop}%
\bibitem [{\citenamefont {Behncke}\ \emph {et~al.}(1979)\citenamefont
  {Behncke}, \citenamefont {Armbruster}, \citenamefont {Folkmann},
  \citenamefont {Hagmann}, \citenamefont {Macdonald},\ and\ \citenamefont
  {Mokler}}]{beh79}%
  \BibitemOpen
  \bibfield  {author} {\bibinfo {author} {\bibfnamefont {H.~H.}\ \bibnamefont
  {Behncke}}, \bibinfo {author} {\bibfnamefont {P.}~\bibnamefont {Armbruster}},
  \bibinfo {author} {\bibfnamefont {F.}~\bibnamefont {Folkmann}}, \bibinfo
  {author} {\bibfnamefont {S.}~\bibnamefont {Hagmann}}, \bibinfo {author}
  {\bibfnamefont {J.~R.}\ \bibnamefont {Macdonald}}, \ and\ \bibinfo {author}
  {\bibfnamefont {P.~H.}\ \bibnamefont {Mokler}},\ }\href {\doibase
  10.1007/BF01409384} {\bibfield  {journal} {\bibinfo  {journal} {Z. Phys. A}\
  }\textbf {\bibinfo {volume} {289}},\ \bibinfo {pages} {333} (\bibinfo {year}
  {1979})}\BibitemShut {NoStop}%
\bibitem [{\citenamefont {Viola}\ \emph {et~al.}(1985)\citenamefont {Viola},
  \citenamefont {Kwiatkowski},\ and\ \citenamefont {Walker}}]{vio85}%
  \BibitemOpen
  \bibfield  {author} {\bibinfo {author} {\bibfnamefont {V.~E.}\ \bibnamefont
  {Viola}}, \bibinfo {author} {\bibfnamefont {K.}~\bibnamefont {Kwiatkowski}},
  \ and\ \bibinfo {author} {\bibfnamefont {M.}~\bibnamefont {Walker}},\ }\href
  {\doibase 10.1103/PhysRevC.31.1550} {\bibfield  {journal} {\bibinfo
  {journal} {Phys. Rev. C}\ }\textbf {\bibinfo {volume} {31}},\ \bibinfo
  {pages} {1550} (\bibinfo {year} {1985})}\BibitemShut {NoStop}%
\bibitem [{\citenamefont {Reinhard}\ \emph {et~al.}(1985)\citenamefont
  {Reinhard}, \citenamefont {Greiner}, \citenamefont {Greenbergand},\ and\
  \citenamefont {Vincent}}]{rei85}%
  \BibitemOpen
  \bibfield  {author} {\bibinfo {author} {\bibfnamefont {J.}~\bibnamefont
  {Reinhard}}, \bibinfo {author} {\bibfnamefont {W.}~\bibnamefont {Greiner}},
  \bibinfo {author} {\bibfnamefont {J.}~\bibnamefont {Greenbergand}}, \ and\
  \bibinfo {author} {\bibfnamefont {P.}~\bibnamefont {Vincent}},\ }\href@noop
  {} {\emph {\bibinfo {title} {Treatise on Heavy-Ion Science}}},\ edited by\
  \bibinfo {editor} {\bibfnamefont {A.}~\bibnamefont {Bromley}},\ Vol.~\bibinfo
  {volume} {5}\ (\bibinfo  {publisher} {Plenum, New York},\ \bibinfo {year}
  {1985})\BibitemShut {NoStop}%
\bibitem [{\citenamefont {Carlson}\ \emph {et~al.}(1968)\citenamefont
  {Carlson}, \citenamefont {Nestor}, \citenamefont {Tucker},\ and\
  \citenamefont {Malik}}]{car68}%
  \BibitemOpen
  \bibfield  {author} {\bibinfo {author} {\bibfnamefont {T.~A.}\ \bibnamefont
  {Carlson}}, \bibinfo {author} {\bibfnamefont {C.~W.}\ \bibnamefont {Nestor}},
  \bibinfo {author} {\bibfnamefont {T.~C.}\ \bibnamefont {Tucker}}, \ and\
  \bibinfo {author} {\bibfnamefont {F.~B.}\ \bibnamefont {Malik}},\ }\href
  {\doibase 10.1103/PhysRev.169.27} {\bibfield  {journal} {\bibinfo  {journal}
  {Phys. Rev.}\ }\textbf {\bibinfo {volume} {169}},\ \bibinfo {pages} {27}
  (\bibinfo {year} {1968})}\BibitemShut {NoStop}%
\bibitem [{\citenamefont {Anholt}(1985)}]{anh85}%
  \BibitemOpen
  \bibfield  {author} {\bibinfo {author} {\bibfnamefont {R.}~\bibnamefont
  {Anholt}},\ }\href {\doibase 10.1103/RevModPhys.57.995} {\bibfield  {journal}
  {\bibinfo  {journal} {Rev. Mod. Phys.}\ }\textbf {\bibinfo {volume} {57}},\
  \bibinfo {pages} {995} (\bibinfo {year} {1985})}\BibitemShut {NoStop}%
\bibitem [{\citenamefont {extracted from~the NuDat 2~database}()}]{nudat}%
  \BibitemOpen
  \bibfield  {author} {\bibinfo {author} {\bibfnamefont {NUDAT2,~National Nuclear Data Center,~Information}\
  \bibnamefont {extracted from~the NuDat 2~database}},\ }\href
  {http://www.nndc.bnl.gov/nudat2/} {\bibinfo  {journal}
  {http://www.nndc.bnl.gov/nudat2/}\ }\BibitemShut {NoStop}%
\bibitem [{\citenamefont {Kib{\'e}di}\ \emph {et~al.}(2008)\citenamefont
  {Kib{\'e}di}, \citenamefont {Burrows}, \citenamefont {Trzhaskovskaya},
  \citenamefont {Davidson},\ and\ \citenamefont {Jr.}}]{Kib08}%
  \BibitemOpen
  \bibfield  {author} {\bibinfo {author} {\bibfnamefont {T.}~\bibnamefont
  {Kib{\'e}di}}, \bibinfo {author} {\bibfnamefont {T.}~\bibnamefont {Burrows}},
  \bibinfo {author} {\bibfnamefont {M.}~\bibnamefont {Trzhaskovskaya}},
  \bibinfo {author} {\bibfnamefont {P.}~\bibnamefont {Davidson}}, \ and\
  \bibinfo {author} {\bibfnamefont {C.~N.}\ \bibnamefont {Jr.}},\ }\href
  {\doibase http://dx.doi.org/10.1016/j.nima.2008.02.051} {\bibfield  {journal}
  {\bibinfo  {journal} {NIM A}\
  }\textbf {\bibinfo {volume} {589}},\ \bibinfo {pages} {202} (\bibinfo {year}
  {2008})}\BibitemShut {NoStop}%
\bibitem [{\citenamefont {Aritomo}(2009)}]{ari09}%
  \BibitemOpen
  \bibfield  {author} {\bibinfo {author} {\bibfnamefont {Y.}~\bibnamefont
  {Aritomo}},\ }\href {\doibase 10.1103/PhysRevC.80.064604} {\bibfield
  {journal} {\bibinfo  {journal} {Phys. Rev. C}\ }\textbf {\bibinfo {volume}
  {80}},\ \bibinfo {pages} {064604} (\bibinfo {year} {2009})}\BibitemShut
  {NoStop}%
\bibitem [{\citenamefont {Shi}\ and\ \citenamefont
  {Danielewicz}(2003)}]{shi03}%
  \BibitemOpen
  \bibfield  {author} {\bibinfo {author} {\bibfnamefont {L.}~\bibnamefont
  {Shi}}\ and\ \bibinfo {author} {\bibfnamefont {P.}~\bibnamefont
  {Danielewicz}},\ }\href {\doibase 10.1103/PhysRevC.68.064604} {\bibfield
  {journal} {\bibinfo  {journal} {Phys. Rev. C}\ }\textbf {\bibinfo {volume}
  {68}},\ \bibinfo {pages} {064604} (\bibinfo {year} {2003})}\BibitemShut
  {NoStop}%
\bibitem [{\citenamefont {Chen}\ \emph {et~al.}(2005)\citenamefont {Chen},
  \citenamefont {Ko},\ and\ \citenamefont {Li}}]{lie05}%
  \BibitemOpen
  \bibfield  {author} {\bibinfo {author} {\bibfnamefont {L.-W.}\ \bibnamefont
  {Chen}}, \bibinfo {author} {\bibfnamefont {C.~M.}\ \bibnamefont {Ko}}, \ and\
  \bibinfo {author} {\bibfnamefont {B.-A.}\ \bibnamefont {Li}},\ }\href
  {\doibase 10.1103/PhysRevLett.94.032701} {\bibfield  {journal} {\bibinfo
  {journal} {Phys. Rev. Lett.}\ }\textbf {\bibinfo {volume} {94}},\ \bibinfo
  {pages} {032701} (\bibinfo {year} {2005})}\BibitemShut {NoStop}%
\bibitem [{\citenamefont {Baran}\ \emph {et~al.}(2005)\citenamefont {Baran},
  \citenamefont {Colonna}, \citenamefont {Toro}, \citenamefont
  {Zielinska-Pfab\'e},\ and\ \citenamefont {Wolter}}]{bar05}%
  \BibitemOpen
  \bibfield  {author} {\bibinfo {author} {\bibfnamefont {V.}~\bibnamefont
  {Baran}}, \bibinfo {author} {\bibfnamefont {M.}~\bibnamefont {Colonna}},
  \bibinfo {author} {\bibfnamefont {M.~D.}\ \bibnamefont {Toro}}, \bibinfo
  {author} {\bibfnamefont {M.}~\bibnamefont {Zielinska-Pfab\'e}}, \ and\
  \bibinfo {author} {\bibfnamefont {H.~H.}\ \bibnamefont {Wolter}},\ }\href
  {\doibase 10.1103/PhysRevC.72.064620} {\bibfield  {journal} {\bibinfo
  {journal} {Phys. Rev. C}\ }\textbf {\bibinfo {volume} {72}},\ \bibinfo
  {pages} {064620} (\bibinfo {year} {2005})}\BibitemShut {NoStop}%
\bibitem [{\citenamefont {Supmat}()}]{supmat}%
  \BibitemOpen
  \bibfield  {author} {\bibinfo {author} {\bibnamefont {See Supplemental Material for TDHF calculations}}\ }\href
  {Supplemental Material} {}\BibitemShut {NoStop}%
\bibitem [{\citenamefont {Simenel}\ \emph {et~al.}(2001)\citenamefont
  {Simenel}, \citenamefont {Chomaz},\ and\ \citenamefont {de~France}}]{sim01}%
  \BibitemOpen
  \bibfield  {author} {\bibinfo {author} {\bibfnamefont {C.}~\bibnamefont
  {Simenel}}, \bibinfo {author} {\bibfnamefont {P.}~\bibnamefont {Chomaz}}, \
  and\ \bibinfo {author} {\bibfnamefont {G.}~\bibnamefont {de~France}},\ }\href
  {http://search.ebscohost.com.in2p3.bib.cnrs.fr/login.aspx?direct=true&db=inh&AN=6904485&lang=fr&site=ehost-live}
  {\bibfield  {journal} {\bibinfo  {journal} {Physical Review Letters}\
  }\textbf {\bibinfo {volume} {86}},\ \bibinfo {pages} {2971 } (\bibinfo {year}
  {2001})}\BibitemShut {NoStop}%
\bibitem [{\citenamefont {Simenel}\ \emph {et~al.}(2007)\citenamefont
  {Simenel}, \citenamefont {Chomaz},\ and\ \citenamefont {de~France}}]{sim07}%
  \BibitemOpen
  \bibfield  {author} {\bibinfo {author} {\bibfnamefont {C.}~\bibnamefont
  {Simenel}}, \bibinfo {author} {\bibfnamefont {P.}~\bibnamefont {Chomaz}}, \
  and\ \bibinfo {author} {\bibfnamefont {G.}~\bibnamefont {de~France}},\ }\href
  {http://search.ebscohost.com.in2p3.bib.cnrs.fr/login.aspx?direct=true&db=inh&AN=9956544&lang=fr&site=ehost-live}
  {\bibfield  {journal} {\bibinfo  {journal} {Physical Review C (Nuclear
  Physics)}\ }\textbf {\bibinfo {volume} {76}},\ \bibinfo {pages} {024609 }
  (\bibinfo {year} {2007})}\BibitemShut {NoStop}%
\bibitem [{\citenamefont {Ka-Hae}\ \emph {et~al.}(1997)\citenamefont {Ka-Hae},
  \citenamefont {Otsuka},\ and\ \citenamefont {Bonche}}]{kim97}%
  \BibitemOpen
  \bibfield  {author} {\bibinfo {author} {\bibfnamefont {K.}~\bibnamefont
  {Ka-Hae}}, \bibinfo {author} {\bibfnamefont {T.}~\bibnamefont {Otsuka}}, \
  and\ \bibinfo {author} {\bibfnamefont {P.}~\bibnamefont {Bonche}},\ }in\
  \href
  {http://search.ebscohost.com.in2p3.bib.cnrs.fr/login.aspx?direct=true&db=inh&AN=5735882&lang=fr&site=ehost-live}
  {\emph {\bibinfo {booktitle} {Journal of Physics G (Nuclear and Particle
  Physics)}}},\ Vol.~\bibinfo {volume} {23}\ (\bibinfo {address} {Dept. of
  Phys., Tokyo Univ., Japan},\ \bibinfo {year} {1997})\ pp.\ \bibinfo {pages}
  {1267 -- 1273}\BibitemShut {NoStop}%
\end{thebibliography}
